\documentclass[journal]{IEEEtran}
\IEEEoverridecommandlockouts
\usepackage{amsmath,amssymb,amsfonts,amsthm}
\usepackage[ruled,linesnumbered,lined]{algorithm2e}
\usepackage{algorithmic}
\usepackage{array}
\usepackage[caption=false,font=footnotesize]{subfig}
\usepackage{textcomp}
\usepackage{stfloats}
\usepackage{url}
\usepackage{verbatim}
\usepackage{graphicx}
\usepackage{cite}
\usepackage{bm}
\usepackage{color, multirow}
\usepackage[table]{xcolor}
\usepackage{colortbl}
\usepackage[colorlinks, linkcolor=blue, citecolor=blue]{hyperref}
\newtheorem{lemma}{Lemma}
\newtheorem{remark}{Remark}

\newtheorem{definition}{Definition}
\newtheorem{proposition}{Proposition}
\DeclareMathOperator*{\argmin}{arg\,min}
\DeclareMathOperator*{\argmax}{arg\,max}

\begin{document}

\title{Transmitter-Side Beyond-Diagonal RIS-Enabled Integrated Sensing and Communications
}
\author{Kexin Chen, Yijie Mao,~\IEEEmembership{Member,~IEEE,} and Wonjae Shin,~\IEEEmembership{Senior Member,~IEEE}
\thanks{A preliminary version of this work was presented at the IEEE Int. Workshop Signal Process. Adv. Wireless Commun. (SPAWC), 2024 \cite{ckx24_transmitterBDRIS}.}
\thanks{This work has been supported in part by the National Nature Science Foundation of China under Grant 62571331. (\textit{Corresponding author: Yijie Mao})}
\thanks{K. Chen and Y. Mao are with the School of Information Science and Technology, ShanghaiTech University, Shanghai 201210, China (email: \{chenkx2023, maoyj\}@shanghaitech.edu.cn).}
\thanks{W. Shin is with the School of Electrical Engineering, Korea University, Seoul 02841, South Korea (email: wjshin@korea.ac.kr).}
\vspace{-0.5cm}
}
\maketitle
\begin{abstract}
Beyond diagonal reconfigurable intelligent surfaces (BD-RIS) have emerged as a promising technology for 6G wireless networks, offering more advanced control over electromagnetic wave propagation than conventional diagonal RIS. This paper proposes a novel integrated sensing and communication (ISAC) framework that incorporates BD-RIS at the transmitter. 
This not only opens the door to enhanced sensing and communication performance, but also alleviates the need for large-scale fully digital radio frequency (RF) chains at the transmitter.
Based on the proposed system model, we formulate a normalized weighted optimization problem to jointly design the active beamforming and the BD-RIS scattering matrix with the aim of jointly minimizing the trace of the Cram{\'e}r-Rao bound (CRB) for sensing targets and maximizing the sum rate (SR) for communication users.
To address this highly coupled optimization problem, we propose a novel and low-complexity iterative algorithm that efficiently solves the active beamforming and scattering matrix subproblems by transforming each into a series of tractable projection problems with closed-form solutions.
Numerical results show the appealing capability of the transmitter-side BD-RIS-aided ISAC over conventional diagonal RIS-aided ISAC in enhancing both sensing and communication performance.
Moreover, compared to the classic iterative algorithm, the proposed algorithm offers enhanced dual-functional performance while significantly reducing the computational complexity.
\end{abstract}

\begin{IEEEkeywords}
Integrated sensing and communication (ISAC), beyond-diagonal reconfigurable intelligent surface (BD-RIS), Cram\'er-Rao bound (CRB).
\end{IEEEkeywords}

\section{Introduction}\label{sec:1_intro}
Driven by the stringent requirements for high-accuracy sensing and ultra-reliable communication in emerging applications such as intelligent transportation and smart cities, next-generation wireless networks are undergoing a paradigm shift toward integrated sensing and communication (ISAC) \cite{lf22_tutorial}.
Recently, the Radio Communication Division of the International Telecommunication Union (ITU-R) has identified ISAC as one of its key usage scenarios for International Mobile Telecommunication 2030 (IMT-2030/6G), thanks to its capability to exploit the potential of wireless networks \cite{kaushik24_ISAC}.
By enabling spectrum sharing and joint signal processing within a same hardware platform, ISAC significantly improves spectral efficiency (SE) and energy efficiency (EE) while reducing hardware complexity and resource rivalry \cite{lf20_ISAC,lf22_tutorial}. 
Moreover, the tighter interplay between sensing and communication in ISAC systems gives rise to coordination and integration gains, thereby facilitating superior performance compared to conventional single-functionality systems \cite{lsh24_ISACtutorial}.
\par
Meanwhile, reconfigurable intelligent surfaces (RIS) have also emerged as a highly promising technology for 6G \cite{lr23_RISISAC}. 
RIS is typically implemented as a two-dimensional array comprising numerous passive elements, each capable of independently adjusting the amplitude and phase of incident waves.
Generally, the deployment of RIS in wireless networks can be categorized into two distinct types:
\textit{serving as part of the wireless channel} \cite{lr23_RISISAC2,wzl23_STARS,wwj24_normalized,ly24_mmWaveISAC,wl25_RISmmWaveISAC}, or \textit{being integrated into the transmitter architecture} \cite{jamali21_activeantenna,buzzi21_transmitRIS,dwn23_ITScommunication}, each offering unique advantages for ISAC systems.
Specifically, in the former case, the RIS is positioned at the side of users and targets, which reconfigures wireless propagation paths and then offers the following advantages: \textit{i)} enhancing the signal strength and suppressing the mutual interference; \textit{ii)} establishing a virtual link for users and targets in shadowed areas, which thereby extends the system coverage and improves the overall dual-functional performance \cite{lr23_RISISAC,lr23_RISISAC2,wzl23_STARS,wwj24_normalized,ly24_mmWaveISAC,wl25_RISmmWaveISAC}.
In contrast, for the latter case, RIS is incorporated into the transmitter architecture to facilitate signal transmission, which yields the following advantages: \textit{i)} alleviating the need for large-scale antennas along with extensive radio frequency (RF) processing at the transmitter, since RIS is capable to reconfigure the phase of incident signals with ultra-low power consumption \cite{jamali21_activeantenna,buzzi21_transmitRIS,dwn23_ITScommunication};
\textit{ii)} eliminating the need for channel estimation between the RIS and active antennas, where the channel between them can be properly designed and remains fixed due to the short distance \cite{jamali21_activeantenna}.
In view of the appealing benefits introduced by RIS, considerable efforts have been dedicated to various RIS-assisted ISAC systems \cite{lr23_RISISAC2,wzl23_STARS, wwj24_normalized, ly24_mmWaveISAC, wl25_RISmmWaveISAC}.
However, most existing ISAC studies only consider the deployment of RIS as part of the wireless channel. The potential of transmitter-side RIS in ISAC is overlooked, despite its investigation in communication-only scenarios
\cite{jamali21_activeantenna,buzzi21_transmitRIS,dwn23_ITScommunication}.
\par
Recently, a revolutionary RIS architecture known as beyond-diagonal RIS (BD-RIS) has emerged \cite{ssp22_BDRIS, lhy22_BDRIS, lhy25_BDRIStutorial}. 
Different from traditional diagonal RIS, where each element is individually controlled, BD-RIS introduces inter-element interactions and thus enables joint processing of the incident signals.
Generally, three BD-RIS architectures have been widely studied: single-, fully-, and group-connected topologies, each offering distinct trade-offs between hardware complexity and system capability \cite{lhy22_BDRIS}.
To be specific, the single-connected topology corresponds to the traditional diagonal RIS.
In contrast, the fully-connected BD-RIS enables interconnections among all elements, offering the highest design flexibility albeit at the cost of increased complexity.
The group-connected architecture generalizes these two extreme cases by partitioning elements into internally fully connected and mutually independent groups.
The additional degrees of freedom (DoF) of BD-RIS has drawn growing attention to its application in ISAC.
In particular, prior works primarily focused on the fully-connected BD-RIS architecture, covering a range of problems such as sum-rate maximization \cite{lzr24_ISAC_BDRIS}, power minimization \cite{gzh24_BDRISISAC}, discrete-valued phase design \cite{esmaeil24_BDRISISAC}, and statistical analysis \cite{Nguyen25_BDRIS_ISAC}.
%
%
More recently, growing efforts have been devoted to exploring other BD-RIS architectures for ISAC, specifically the group-connected BD-RIS \cite{pxl24_BDRIS_ISAC} and the cell-wise BD-RIS with hybrid transmitting-reflecting capability \cite{WBW23_BDRISISAC}.
%
It is also worth noting that beyond sensing metrics such as the signal-to-noise ratio (SNR) utilized in \cite{lzr24_ISAC_BDRIS,gzh24_BDRISISAC,esmaeil24_BDRISISAC,Nguyen25_BDRIS_ISAC}, beampattern gain in \cite{pxl24_BDRIS_ISAC} and signal-to-clutter-plus-noise ratio (SCNR) in \cite{WBW23_BDRISISAC}, the Cram\'er-Rao bound (CRB) considered in \cite{zs25_BDRIS_ISAC, wdw25_BDRIS_ISAC,zxq25_BDRISISAC} is actually the widely recognized fundamental metric, which characterizes the theoretical limit of parameter estimation accuracy \cite{lj08_CRB}.
All these studies highlight the superiority of BD-RIS in enhancing the dual-functional performance compared to conventional diagonal RIS.
However, the focus of existing works remains on deploying BD-RIS as part of the wireless channel, and the potential benefits of incorporating BD-RIS into the ISAC transmitter has not been investigated yet.
\par
From an algorithmic perspective, the deployment of BD-RIS imposes significant challenges to ISAC beamforming design compared to scenarios without RIS or those aided by traditional diagonal RIS. 
The difficulty typically arises from: 
\textit{i)} the strong coupling among variables such as the active beamforming and BD-RIS scattering matrix in communication and sensing metrics; 
\textit{ii)} the non-convex symmetric and orthogonality constraint inherent in the structure of BD-RIS.
It is also worth noting that when choosing the well-known sensing CRB for multi-parameter estimation, the resulting blockwise matrix structure often introduces additional complexity to the optimization problem \cite{lj08_CRB}.
To address this beamforming issue, one common method is to leverage conventional iterative algorithms.
Specifically, variable decoupling is realized by transforming the original non-convex problem into more tractable subproblems. This is achieved via approaches such as alternating optimization (AO) \cite{pxl24_BDRIS_ISAC, wdw25_BDRIS_ISAC,zxq25_BDRISISAC}, block coordinate descent (BCD) \cite{gzh24_BDRISISAC, lzr24_ISAC_BDRIS}, or alternating direction method of multipliers (ADMM) \cite{WBW23_BDRISISAC}. 
Subsequently, to iteratively solve the resulting subproblems with respect to different variables, aforementioned strategies are typically combined with methods such as weighted minimum mean squared error (WMMSE) \cite{pxl24_BDRIS_ISAC,zxq25_BDRISISAC}, majorization minimization (MM) \cite{lzr24_ISAC_BDRIS,gzh24_BDRISISAC}, manifold optimization \cite{pxl24_BDRIS_ISAC}, semidefinite relaxation (SDR) \cite{wdw25_BDRIS_ISAC}, or penalty dual decomposition (PDD) \cite{lzr24_ISAC_BDRIS,gzh24_BDRISISAC,wdw25_BDRIS_ISAC, zs25_BDRIS_ISAC}.
In other words, the tractable subproblems are typically solved via a hybrid use of various iterative methods, with the resulting convex subproblems handled by solvers like CVX.
However, the iterative use of convex solvers in these studies inevitably results in high computational complexity, hindering its applications in large-scale networks.
\par
In summary, although a growing body of studies \cite{lzr24_ISAC_BDRIS,gzh24_BDRISISAC,esmaeil24_BDRISISAC,Nguyen25_BDRIS_ISAC,pxl24_BDRIS_ISAC,WBW23_BDRISISAC,zs25_BDRIS_ISAC,wdw25_BDRIS_ISAC,zxq25_BDRISISAC} have explored the incorporation of BD-RIS into ISAC, several key challenges and limitations remain unsolved.
\textit{Firstly,} the scope of current research on BD-RIS-aided ISAC is limited to user-side BD-RIS. 
The potential of transmitter-side RIS has been largely overlooked in ISAC systems, despite its demonstrated benefits in communication systems with either diagonal \cite{jamali21_activeantenna,buzzi21_transmitRIS,dwn23_ITScommunication} or beyond-diagonal \cite{Mishra23_BDRIScomm} structures.
\textit{Secondly,} most existing studies focus on optimizing communication performance under sensing constraints or vice versa \cite{wzl23_STARS, ly24_mmWaveISAC, wl25_RISmmWaveISAC,lzr24_ISAC_BDRIS,gzh24_BDRISISAC,Nguyen25_BDRIS_ISAC,pxl24_BDRIS_ISAC,WBW23_BDRISISAC,zs25_BDRIS_ISAC,wdw25_BDRIS_ISAC,zxq25_BDRISISAC}, leaving the trade-off between the two in RIS-assisted ISAC systems largely unexplored. This gap is primarily due to the strongly non-convex formulations induced by their joint optimization.
\textit{Thirdly,} the reliance on highly complex optimization algorithms limits their application to large-scale networks.
Motivated by the above discussion, we initiate the study of a transmitter-side BD-RIS-aided ISAC system, where an efficient AO-based algorithm is proposed to solve the multi-objective problem.
The main contributions of this paper are outlined as follows:
\begin{itemize}
\item We propose a novel BD-RIS-enabled transmitter architecture for ISAC systems, where the BD-RIS is placed a few wavelengths away from the active antennas to fully exploit the additional DoFs of BD-RIS.
To the best of our knowledge, this is the first work to explore the potential of transmitter-side BD-RIS in ISAC, as all previous studies have considered only user-side BD-RISs.
\item To strike a balance between sensing and communication, we formulate a normalized weighted optimization problem, where the active beamforming and the BD-RIS scattering matrix are optimized to jointly maximize the normalized communication sum rate (SR) and minimize the normalized trace of sensing CRB.
We propose a highly efficient and novel AO framework to iteratively optimize the two highly coupled variables.
To reduce computational complexity, each AO subproblem is solved using a novel projected successive convex approximation (PSCA) method, which transforms it into a sequence of tractable projection problems with closed-form solutions.
Notably, the proposed algorithm is general since it can be directly applied to conventional joint active and passive beamforming problems in user-side diagonal RIS-assisted ISAC, while achieving significantly lower computational complexity than conventional approaches.
\item Extensive numerical results are presented to show the efficacy of the proposed framework and algorithm. 
It is observed that the proposed transmitter-side BD-RIS-aided framework offers superior communication and sensing performance over the conventional model aided by diagonal RIS. 
Moreover, the proposed algorithm achieves the same performance as the classic WMMSE-PDD-based algorithm, while substantially reducing computational complexity.
\end{itemize}
\par
\textit{Organizations:} The remainder of this paper is structured as follows. 
Section \ref{sec:2_model} presents the system model and problem formulation. 
In Section \ref{sec:3_algor}, the weighted optimization problem is solved by the proposed algorithm.
Section \ref{sec:4_result} shows the numerical results, followed by a conclusion in Section \ref{sec:5_conclu}.
\par
\textit{Notations:} Matrices and vectors are denoted by the boldface upper and lower case letters, respectively. 
$(\cdot)^T$, $(\cdot)^\ast$, $(\cdot)^H$, $\mathrm{tr}(\cdot)$ and $(\cdot)^{-1}$ refer to the transpose, conjugate, conjugate transpose, trace and inverse operators, respectively. 
$\mathrm{diag}(\mathbf{a})$ denotes a diagonal matrix with $\mathbf{a}$ on its main diagonal, while $\mathrm{blkdiag}(\mathbf{A}_1,\dots,\mathbf{A}_L)$ stands for a block-diagonal matrix involving blocks $\mathbf{A}_1,\dots,\mathbf{A}_L$ on its diagonal. 
The all-one and all-zero matrices of dimension $L\times P$ are denoted by $\mathbf{1}_{L\times P}$, and $\mathbf{0}_{L\times P}$, respectively. 
$\mathrm{Re}(\cdot)$ and $\mathrm{Im}(\cdot)$ refer to the real and imaginary component of a complex, respectively. $\odot$ represents the hadamard product, and $\langle\cdot,\cdot\rangle$ denotes the inner product in the complex matrix space.
\section{System Model and Problem Formulation} \label{sec:2_model}
\begin{figure}
    \centering
    \includegraphics[width=0.97\linewidth]{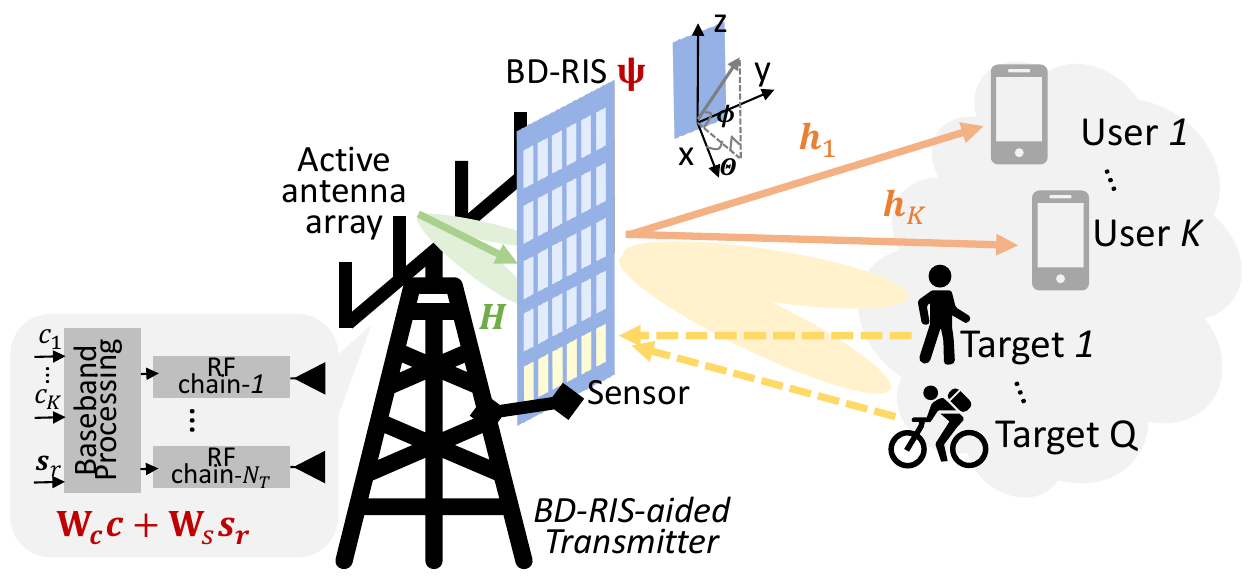}
    \caption{The system model of BD-RIS-aided transmitter architecture for ISAC.}
    \label{fig: model}
\end{figure}

As illustrated in Fig. \ref{fig: model}, we consider a multi-user multi-target ISAC system over a coherent processing interval (CPI) of length $M$. The BD-RIS-aided transmitter seeks to estimate the parameters of $Q$ point-like sensing targets indexed by $\mathcal{Q}\triangleq\left \{ 1,\dots, Q \right \}$. Simultaneously, it intends to transmit data to $K$ single-antenna communication users indexed by $\mathcal{K}\triangleq\left\{1,\dots, K \right \}$. 
\par
At the transmitter, the dual-functional signal is first processed via $N_T$ RF chains, each connected to a dedicated active feed antenna, and then transmitted directly to a BD-RIS with $N_I$ elements. The corresponding scattering matrix of the BD-RIS is $\bm{\Psi} \in \mathbb{C}^{N_I \times N_I}$.
It is assumed that the BD-RIS is positioned a short distance, typically a few wavelengths away from the active antennas. 
In this work, we consider a BD-RIS operating in the transmissive mode\footnote{This choice is motivated by its demonstrated advantage of reduced feed blockage when placed a few wavelengths away from the active antennas, in contrast to its reflective counterpart \cite{jamali21_activeantenna}.}, which allows incident signals to penetrate it only \cite{lhy22_BDRIS}. 
Moreover, by deploying a low-cost sensor with $N_S$ elements at the BD-RIS \cite{wzl23_STARS,sxd22_sensingRIS,wwj24_normalized}, the received echo signals experience a three-hop link, i.e., active antennas at the transmitter$\to$BD-RIS$\to$targets$\to$sensor.  
This setup leads to reduced number of hops and path loss compared to conventional user-side RIS-assisted systems, where sensing is performed at a receiver located far from the RIS.
It is also assumed that the antenna-facing side of the sensor is physically blocked to ensure unidirectional reception of echo signals without self-interference from transmit signals \cite{wzl23_STARS}.
\subsection{BD-RIS Architecture}\label{sec:2_A}
Following \cite{ssp22_BDRIS, lhy22_BDRIS}, we model the BD-RIS as a lossless and reciprocal $N_I$-port reconfigurable impedance network, where the scattering matrix $\bm{\Psi}$ is determined by the underlying circuit topology. Three classical BD-RIS architectures are considered in this work, namely, single-, fully-, and group-connected configurations, each is mathematically modeled as:
\subsubsection{Single-connected BD-RIS} In this architecture, each port of BD-RIS is connected to the ground via a reconfigurable impedance, which is independent of any other ports. The scattering matrix $\bm{\Psi}$ is therefore exhibiting a diagonal structure satisfying
\begin{equation}
    \mathcal{M}_1\triangleq\{\bm{\Psi}|\bm{\Psi}=\mathrm{diag}(\psi_{1},\dots,\psi_{N_I})\},
\end{equation}
where $\psi_{i}$ denotes the transmission coefficients satisfying $|\psi_{i}|=1, \forall i\in\{1,\dots,N_I\}$.
\subsubsection{Fully-connected BD-RIS} This topology realizes full connectivity among ports, where any pairs of them are interconnected via reconfigurable impedances. Hence, its scattering matrix $\bm{\Psi}$ is a full matrix satisfying the following symmetric and unitary constraint
\begin{equation}
\mathcal{M}_2\triangleq\{\bm{\Psi}|\bm{\Psi}=\bm{\Psi}^T, \bm{\Psi}^H\bm{\Psi}=\mathbf{I}_{N_I}\}.
\end{equation}
\subsubsection{Group-connected BD-RIS} This architecture is proposed to achieve trade-offs between the hardware complexity and system capability. Specifically, the $N_I$ ports are divided into $G$ groups, where the ports within the same group are fully-connected, and those in different groups are independent of each other. Let $\mathcal{G}=\{1,\dots,G\}$ denote the set of the group index, and $N_{I,g}, \forall g\in\mathcal{G}$ denote the number of ports in the $g$-th group satisfying $\sum_{g\in\mathcal{G}} N_{I,g} = N_I$. The scattering matrix $\bm{\Psi}$ is modeled as
\begin{equation}
\begin{aligned}
    \mathcal{M}_3\triangleq\{\bm{\Psi}|&\bm{\Psi}=\mathrm{blkdiag}(\bm{\Psi}_1,\dots,\bm{\Psi}_G),\\
    &\bm{\Psi}_g=\bm{\Psi}_g^T, \bm{\Psi}_g^H\bm{\Psi}_g=\mathbf{I}_{N_{I,g}}, \forall g \in \mathcal{G}\},
\end{aligned}
\end{equation}
where $\bm{\Psi}_g, \forall g$ are complex symmetric unitary matrices.
\par
It is obvious that the group-connected topology bridges the other two extreme cases. Specifically, when $G=1$, it becomes the fully-connected BD-RIS with the highest design flexibility, while with $G=N_I$, it boils down to the single-connected BD-RIS with the lowest hardware complexity.
\subsection{Signal Model and Performance Metrics}
At the transmitter, the dual-functional signal is processed for simultaneous communication with $K$ users and sensing of $Q$ targets. 
Let $\mathbf{c}[m]\triangleq\left[c_{1}[m],\dots,c_{K}[m]\right]^{T}\in \mathbb{C}^{K\times 1}$ denote the zero-mean information symbol vector at time index $m$, which satisfies $\mathbb{E}[\mathbf{c}[m]\mathbf{c}^{H}[m]]=\mathbf{I}_{K}$. The corresponding active beamforming matrix is written as $\mathbf{W}_c\triangleq[\mathbf{w}_{1},\dots, \mathbf{w}_{K}]\in \mathbb{C}^{N_{T}\times K}$. 
To ensure promising sensing performance, a dedicated sensing signal $\mathbf{s}_r[m] \in \mathbb{C}^{N_{T}\times 1}$ is considered satisfying $\mathbb{E}[\mathbf{s}_r[m]\mathbf{s}_r^{H}[m]]=\mathbf{I}_{N_T}$ and $\mathbb{E}[\mathbf{c}[m]\mathbf{s}^{H}_r[m]]=\mathbf{0}_{K\times N_T}$.
It is multiplied by the sensing beamforming matrix $\mathbf{W}_s \in \mathbb{C}^{N_T\times N_T}$ before transmission \cite{lzr24_ISAC_BDRIS}. 
The overall active beamforming matrix is $\mathbf{W}\triangleq[\mathbf{W}_c,\mathbf{W}_s] \in \mathbb{C}^{N_T\times (K+N_T)}$, which remains consistent in one CPI.
The transmit signal at time index $m$ is expressed as
\begin{equation}\label{eq:transmit signal}
	\mathbf{x}[m]=\sum_{k\in\mathcal{K}}\mathbf{w}_k c_k[m]+\mathbf{W}_s\mathbf{s}_r[m],
\end{equation}
where the sample covariance matrix of the transmit signal is given by $\mathbf{R}_{x}\triangleq \frac{1}{M}\sum_{m\in \mathcal{M}}\mathbf{x}[m]\mathbf{x}[m]^{H}\approx\mathbb{E}\{\mathbf{x}[m]\mathbf{x}[m]^{H}\}=\mathbf{W}\mathbf{W}^{H}$, with $M$ assumed to be sufficiently large \cite{wwj24_normalized}. The total power budget at the transmitter is $P$, i.e., $\mathrm{tr}(\mathbf{W}\mathbf{W}^{H})\leq P$. 
\par
The transmit signal is first directed to the BD-RIS, which is then forwarded towards the $K$ communication users and $Q$ sensing targets. 
Each element of the channel matrix $\mathbf{H} \in \mathbb{C}^{N_I \times N_T}$ between the active antennas and the BD-RIS is modeled as \cite{jamali21_activeantenna,Mishra23_BDRIScomm}
\begin{equation}\label{eq:HT}
    \begin{aligned}
    \mathbf{H}(i,j)=\frac{\lambda\sqrt{\rho_{\mathrm{eff}} G^A(\theta^\circ_{i,j},\phi^\circ_{i,j}) G^P(\theta^\circ_{i,j},\phi^\circ_{i,j})}}{4 \pi d_{i,j}}e^{-j \frac{2 \pi d_{i,j}}{\lambda}}&,\\
    i\in\{1,\dots,N_I\},\ j\in\{1,\dots,N_T\}&,
    \end{aligned}
\end{equation}
where $\rho_{\mathrm{eff}}\in\mathbb{R}$ denotes the power efficiency of BD-RIS in transmitting the incident signals, while $G^A(\theta^\circ_{i,j},\phi^\circ_{i,j})$ and $G^P(\theta^\circ_{i,j},\phi^\circ_{i,j})$ represent the active and passive antenna gains between the $i$-th BD-RIS element and the $j$-th active antenna with the azimuth and elevation angle $\theta^\circ_{i,j}$, $\phi^\circ_{i,j}$, respectively \cite{jamali21_activeantenna, Mishra23_BDRIScomm, buzzi21_transmitRIS}. $d_{i,j}$ is the distance between the $i$-th BD-RIS element and the $j$-th active antenna, and $\lambda$ is the wavelength of the carrier frequency.
\subsubsection{Communication users}
Let $\mathbf{h}_{k}\in \mathbb{C}^{N_{I}\times 1}$ denote the perfectly known channel vector between the BD-RIS and the $k$-th communication user, and the resulting effective channel between the transmitter and user $k$ is $\mathbf{g}_k^H=\mathbf{h}_{k}^{H}\mathbf{\Psi}\mathbf{H}$. The receive signal of user $k$ at time index $m$ is then given by
\begin{equation} \label{eq:signal_c}
	\begin{aligned}
		y_{k}[m]&=\mathbf{g}_k^H\mathbf{w}_{k}c_{k}[m]+\sum_{i\in \mathcal{K}, i\neq k}\mathbf{g}_k^H \mathbf{w}_{i}c_{i}[m]\\
        &+\mathbf{g}_k^H\mathbf{W}_s\mathbf{s}_r[m]+z_{k}[m], \forall k\in \mathcal{K},
	\end{aligned}
\end{equation}
where $z_{k}[m]$ is the additive white Gaussian noise (AWGN) following $\mathcal{CN}(0,\sigma _{c}^{2})$. 
\par
The signal-to-interference-plus-noise ratio (SINR) for decoding the desired data stream at the $k$-th user is calculated as
\begin{equation}\label{eq:SINR}
    \gamma_k=\frac{\left|\mathbf{g}_{k}^H\mathbf{w}_{k}\right|^2}{\sum_{i\neq k}^K\left|\mathbf{g}_{k}^H \mathbf{w}_{i}\right|^2+\|\mathbf{g}_{k}^H\mathbf{W}_s\|_2^2+\sigma_c^2}, \forall k \in \mathcal{K}.
\end{equation}
In this work, the communication performance is evaluated by the SR metric, which is defined as $R_{\mathrm{sum}}\triangleq\sum_{k\in\mathcal{K}}\log(1+\gamma_k)$.
\subsubsection{Sensing targets}
The dual-functional signal $\mathbf{x}[m]$ is reused to simultaneously fulfill the sensing functionality. At the time index $m$, the echo signal reflected from $Q$ targets and received at the sensor is given by
\begin{equation}
	\mathbf{y}_{s}[m]=\mathbf{B}\mathbf{U}\mathbf{A}^{T}\mathbf{\Psi}\mathbf{H}\mathbf{x}[m]+\mathbf{z}_{s}[m],
\end{equation}
where
\begin{equation} \label{eq:ABU def}
	\begin{aligned}
		&\mathbf{B}\triangleq[\mathbf{b}(\theta _1, \phi_1),\dots,\mathbf{b}(\theta _{Q}, \phi_Q)],\\
		&\mathbf{A}\triangleq[\mathbf{a}(\theta _1, \phi_1),\dots,\mathbf{a}(\theta _{Q}, \phi_Q)],\\
        &\mathbf{U}\triangleq\mathrm{diag}(\bm{\alpha}),\ \bm{\alpha}\triangleq[\alpha _{1},\dots,\alpha _{Q}], 
	\end{aligned}
\end{equation}
$\alpha_{q}, \forall q$ represents the complex reflection coefficient of the $q$-th target, and
$\mathbf{z}_{s}[m] \in \mathbb{C}^{N_{S}\times 1}$ following $\mathcal{CN}(\mathbf{0}_{N_{S}\times 1},\sigma _{s}^{2}\mathbf{I}_{N_{S}})$ is the AWGN at the sensor.
$\theta_q$ and $\phi_q, \forall q$ in $\mathbf{b}(\theta _{q}, \phi_q)$ and $\mathbf{a}(\theta_q, \phi_q)$ refer to the azimuth and elevation angles of target $q$ related to the BD-RIS.
In this work, we assume that the BD-RIS is deployed in the $yz$ plane and the sensor is deployed along the $y$ axis. 
The corresponding Cartesian coordinates are $[\mathbf{0},\bm{r}_y,\bm{r}_z] \in \mathbb{R}^{N_I \times 3}$ and $[\mathbf{0},\hat{\bm{r}}_y,\mathbf{0}] \in \mathbb{R}^{N_S \times 3}$, where $\bm{r}_y, \bm{r}_z\in \mathbb{R}^{N_I \times 1}$ denote the $y$- and $z$-coordinate of the BD-RIS elements, and $\hat{\bm{r}}_y\in \mathbb{R}^{N_S \times 1}$ denotes the $y$-coordinate of the sensor, respectively.  
The transmit and receive steering vectors $\mathbf{a}\in\mathbb{C}^{N_{I}\times 1}$ and $\mathbf{b}\in\mathbb{C}^{N_{S}\times 1}$ are therefore calculated as
\begin{equation}
    \begin{aligned}
        \mathbf{a}(\theta _q, \phi_q)&\triangleq e^{-j\frac{2\pi }{\lambda}(\mathbf{r}_y\sin(\theta_q)\cos(\phi_q)+\mathbf{r}_z \sin(\phi_q))},\\
        \mathbf{b}(\theta _q, \phi_q)&\triangleq e^{-j\frac{2\pi }{\lambda} (\hat{\mathbf{r}}_y\sin(\theta_q)\cos(\phi_q))}, \forall q \in \mathcal{Q}.
    \end{aligned}
\end{equation}
\begin{figure*}[t]
    \centering
\begin{equation}\label{eq:FIM entry}
	\begin{aligned}
		\mathbf{F}_{11}&\triangleq(\dot{\mathbf{B}}_{\bm{\theta}}^H \dot{\mathbf{B}}_{\bm{\theta}})\odot (\mathbf{U}\mathbf{A}^{T}\bm{\Psi}\mathbf{H}\mathbf{R}_{x}\mathbf{H}^H\bm{\Psi}^H\mathbf{A}^\ast\mathbf{U}^H)^T+(\dot{\mathbf{B}}_{\bm{\theta}}^H\mathbf{B})\odot (\mathbf{U}\dot{\mathbf{A}}_{\bm{\theta}}^{T}\bm{\Psi}\mathbf{H}\mathbf{R}_{x}\mathbf{H}^H\bm{\Psi}^H\mathbf{A}^\ast\mathbf{U}^H)^T\\&+(\mathbf{B}^{H}\dot{\mathbf{B}}_{\bm{\theta}})\odot (\mathbf{U}\mathbf{A}^{T}\bm{\Psi}\mathbf{H}\mathbf{R}_{x}\mathbf{H}^H\bm{\Psi}^H\dot{\mathbf{A}}_{\bm{\theta}}^\ast\mathbf{U}^H)^T+(\mathbf{B}^{H}\mathbf{B})\odot (\mathbf{U}\dot{\mathbf{A}}_{\bm{\theta}}^{T}\bm{\Psi}\mathbf{H}\mathbf{R}_{x}\mathbf{H}^H\bm{\Psi}^H\dot{\mathbf{A}}_{\bm{\theta}}^\ast\mathbf{U}^H)^T,\\
        \mathbf{F}_{12}&\triangleq(\dot{\mathbf{B}}_{\bm{\theta}}^H \dot{\mathbf{B}}_{\bm{\phi}})\odot (\mathbf{U}\mathbf{A}^{T}\bm{\Psi}\mathbf{H}\mathbf{R}_{x}\mathbf{H}^H\bm{\Psi}^H\mathbf{A}^\ast\mathbf{U}^H)^T+(\dot{\mathbf{B}}_{\bm{\theta}}^H\mathbf{B})\odot (\mathbf{U}\dot{\mathbf{A}}_{\bm{\phi}}^{T}\bm{\Psi}\mathbf{H}\mathbf{R}_{x}\mathbf{H}^H\bm{\Psi}^H\mathbf{A}^\ast\mathbf{U}^H)^T\\&+(\mathbf{B}^{H}\dot{\mathbf{B}}_{\bm{\phi}})\odot (\mathbf{U}\mathbf{A}^{T}\bm{\Psi}\mathbf{H}\mathbf{R}_{x}\mathbf{H}^H\bm{\Psi}^H\dot{\mathbf{A}}_{\bm{\theta}}^\ast\mathbf{U}^H)^T+(\mathbf{B}^{H}\mathbf{B})\odot (\mathbf{U}\dot{\mathbf{A}}_{\bm{\phi}}^{T}\bm{\Psi}\mathbf{H}\mathbf{R}_{x}\mathbf{H}^H\bm{\Psi}^H\dot{\mathbf{A}}_{\bm{\theta}}^\ast\mathbf{U}^H)^T,\\
		\mathbf{F}_{22}&\triangleq(\dot{\mathbf{B}}_{\bm{\phi}}^H \dot{\mathbf{B}}_{\bm{\phi}})\odot (\mathbf{U}\mathbf{A}^{T}\bm{\Psi}\mathbf{H}\mathbf{R}_{x}\mathbf{H}^H\bm{\Psi}^H\mathbf{A}^\ast\mathbf{U}^H)^T+(\dot{\mathbf{B}}_{\bm{\phi}}^H\mathbf{B})\odot (\mathbf{U}\dot{\mathbf{A}}_{\bm{\phi}}^{T}\bm{\Psi}\mathbf{H}\mathbf{R}_{x}\mathbf{H}^H\bm{\Psi}^H\mathbf{A}^\ast\mathbf{U}^H)^T\\&+(\mathbf{B}^{H}\dot{\mathbf{B}}_{\bm{\phi}})\odot (\mathbf{U}\mathbf{A}^{T}\bm{\Psi}\mathbf{H}\mathbf{R}_{x}\mathbf{H}^H\bm{\Psi}^H\dot{\mathbf{A}}_{\bm{\phi}}^\ast\mathbf{U}^H)^T+
        (\mathbf{B}^{H}\mathbf{B})\odot (\mathbf{U}\dot{\mathbf{A}}_{\bm{\phi}}^{T}\bm{\Psi}\mathbf{H}\mathbf{R}_{x}\mathbf{H}^H\bm{\Psi}^H\dot{\mathbf{A}}_{\bm{\phi}}^\ast\mathbf{U}^H)^T,\\
		\mathbf{F}_{13}&\triangleq(\dot{\mathbf{B}}_{\bm{\theta}}^H \mathbf{B})\odot (\mathbf{A}^{T}\bm{\Psi}\mathbf{H}\mathbf{R}_{x}\mathbf{H}^H\bm{\Psi}^H\mathbf{A}^\ast\mathbf{U}^H)^T+
        (\mathbf{B}^H \mathbf{B})\odot (\mathbf{A}^{T}\bm{\Psi}\mathbf{H}\mathbf{R}_{x}\mathbf{H}^H\bm{\Psi}^H\dot{\mathbf{A}}_{\bm{\theta}}^\ast\mathbf{U}^H)^T,\\
		\mathbf{F}_{23}&\triangleq(\dot{\mathbf{B}}_{\bm{\phi}}^H \mathbf{B})\odot (\mathbf{A}^{T}\bm{\Psi}\mathbf{H}\mathbf{R}_{x}\mathbf{H}^H\bm{\Psi}^H\mathbf{A}^\ast\mathbf{U}^H)^T+
        (\mathbf{B}^H \mathbf{B})\odot (\mathbf{A}^{T}\bm{\Psi}\mathbf{H}\mathbf{R}_{x}\mathbf{H}^H\bm{\Psi}^H\dot{\mathbf{A}}_{\bm{\phi}}^\ast\mathbf{U}^H)^T,\\
		\mathbf{F}_{33}&\triangleq
        (\mathbf{B}^H \mathbf{B})\odot (\mathbf{A}^{T}\bm{\Psi}\mathbf{H}\mathbf{R}_{x}\mathbf{H}^H\bm{\Psi}^H\mathbf{A}^\ast)^T. 
	\end{aligned}
\end{equation}
\rule{\linewidth}{0.5pt}
\end{figure*}
\par
The widely used CRB metric, obtained from the inverse of the Fisher information matrix (FIM), is selected to evaluate the sensing estimation performance \cite{lj08_CRB}.
In this work, we aim to estimate the parameters of the azimuth angle $\theta_q$, the elevation angle $\phi_q$, and the complex reflection coefficient $\alpha_q$ for the $Q$ sensing targets.
The parameter set is defined as $\bm{\xi }\triangleq \left \{ \bm{\theta} ,\bm{\phi}, \mathrm{Re}(\bm{\alpha}), \mathrm{Im}(\bm{\alpha}) \right \}^{T}\in \mathbb{R}^{4Q \times 1}$, where $\bm{\theta}\triangleq [\theta_1,\dots,\theta_Q]$ and $\bm{\phi}\triangleq[\phi_1,\dots,\phi_Q]$.
The FIM $\mathbf{F} \in \mathbb{R}^{4Q \times 4Q}$ associated with the parameter set $\bm{\xi }$ is given by 
\begin{equation}\label{eq:FIM}
\scalebox{0.95}{$
	\mathbf{F}=\frac{2M}{\sigma_s^2}\begin{bmatrix}
		\mathrm{Re}(\mathbf{F}_{11})&\mathrm{Re}(\mathbf{F}_{12})&\mathrm{Re}(\mathbf{F}_{13})&-\mathrm{Im}(\mathbf{F}_{13})\\ 
		\mathrm{Re}(\mathbf{F}^T_{12}) & \mathrm{Re}(\mathbf{F}_{22}) & \mathrm{Re}(\mathbf{F}_{23}) & -\mathrm{Im}(\mathbf{F}_{23})\\ 
		\mathrm{Re}(\mathbf{F}^T_{13}) & \mathrm{Re}(\mathbf{F}^T_{23}) & \mathrm{Re}(\mathbf{F}_{33}) & -\mathrm{Im}(\mathbf{F}_{33})\\ 
		-\mathrm{Im}(\mathbf{F}^T_{13}) & -\mathrm{Im}(\mathbf{F}^T_{23}) & -\mathrm{Im}(\mathbf{F}^T_{33}) & \mathrm{Re}(\mathbf{F}_{33})
	\end{bmatrix},$}
\end{equation}
with block entries shown in \eqref{eq:FIM entry} at the top of next page. In \eqref{eq:FIM entry}, $\mathbf{R}_x$ is the transmit covariance matrix specified below \eqref{eq:transmit signal}, $\mathbf{A}$, $\mathbf{B}$, $\mathbf{U}$ are given in \eqref{eq:ABU def}, and
\begin{equation}
	\begin{aligned}
	&	\dot{\mathbf{A}}_{\bm{\theta}}=\begin{bmatrix}
			\frac{\partial \mathbf{a}(\theta _1,\phi_1)}{\partial \theta _1}&,\dots,&\frac{\partial \mathbf{a}(\theta_Q,\phi_Q)}{\partial \theta _{Q}} 
		\end{bmatrix},\\
        &\dot{\mathbf{A}}_{\bm{\phi}}=\begin{bmatrix}
			\frac{\partial \mathbf{a}(\theta _1,\phi_1)}{\partial \phi _1}&,\dots,&\frac{\partial \mathbf{a}(\theta_Q,\phi_Q)}{\partial \phi _{Q}} 
		\end{bmatrix},\\
		&\dot{\mathbf{B}}_{\bm{\theta}}=\begin{bmatrix}
			\frac{\partial \mathbf{b}(\theta _1,\phi_1)}{\partial \theta _1}&,\dots,&\frac{\partial \mathbf{b}(\theta_Q,\phi_Q)}{\partial \theta _{Q}} 
		\end{bmatrix},\\
        &\dot{\mathbf{B}}_{\bm{\phi}}=\begin{bmatrix}
			\frac{\partial \mathbf{b}(\theta _1,\phi_1)}{\partial \phi _1}&,\dots,&\frac{\partial \mathbf{b}(\theta_Q,\phi_Q)}{\partial \phi _{Q}} 
		\end{bmatrix}.
	\end{aligned}
\end{equation}
The detailed derivation procedure to obtain \eqref{eq:FIM} is provided in Appendix \ref{appendix A}. 
Note that the CRB metric in this work is established based on \cite{wzl23_STARS,fty25_ISAClowcomplexity}, but is more general than the one considered therein.
Specifically, the FIM in \eqref{eq:FIM} captures the estimation of multiple sensing targets enabled by the BD-RIS, where the single-target case in \cite{wzl23_STARS} emerges as a special case. Moreover, unlike \cite{fty25_ISAClowcomplexity}, which considers ISAC systems without RIS, the FIM obtained here explicitly incorporates the impact of the BD-RIS scattering matrix design.
\subsection{Problem Formulation}
To enhance the performance trade-off between multi-user communication and multi-target parameter estimation, we formulate a novel normalized weighted optimization problem for the proposed system model. Specifically, the scattering matrix $\bm{\Psi}$ of the BD-RIS and the active beamforming matrix $\mathbf{W}$ are jointly designed to maximize the communication SR, i.e., $R_{\mathrm{sum}}$, while minimizing the trace of the sensing CRB, i.e., $\mathrm{tr}(\mathbf{F^{-1}})$. 
The optimization problem is formulated as
\begin{equation}\label{p:joint optimize}
\mathop{\max}_{\mathbf{W}\in\mathcal{W},\bm{\Psi}\in\mathcal{M}_i}\,\,\, \rho\frac{R_{\mathrm{sum}}}{V_c}+(1-\rho)\frac{-\mathrm{tr}(\mathbf{F}^{-1})}{V_s},
\end{equation}
where depending on the specific BD-RIS architectures described in Subsection \ref{sec:2_A}, $\mathcal{M}_i$ varies accordingly. This work aims to develop a unified solution framework that is applicable to all BD-RIS architectures, i.e., $\forall i\in\{1, 2, 3\}$. 
Moreover, the weighting factor $\rho\in (0,1)$ is used to enable a flexible trade-off between communication and sensing.
The feasible set $\mathcal{W}$ of the active beamforming $\mathbf{W}$ is defined as $\mathcal{W}\triangleq\{\mathbf{W}|\mathrm{tr}(\mathbf{W}\mathbf{W}^H)=P\}$, since any locally optimal solution to problem \eqref{p:joint optimize} utilizes the full available power \cite{fty25_ISAClowcomplexity}. 
It is also worth noting that to combine the communication and sensing metrics into an overall objective, we normalize $R_{\mathrm{sum}}$ and $\mathrm{tr}(\mathbf{F}^{-1})$ via their upper/lower bound $V_c$ and $V_s$, respectively \cite{wwj24_normalized}. 
Specifically, the communication normalization constant $V_c$ is obtained by maximizing the SR only, which is given as
\begin{equation}\label{p:com-only}
V_c=\argmax_{\mathbf{W}\in\mathcal{W},\bm{\Psi}\in\mathcal{M}_i}\,\,\, R_{\mathrm{sum}}. 
\end{equation}
Additionally, $V_s$ is obtained by minimizing the trace of the sensing CRB, which is expressed as
\begin{equation}\label{p:sen-only}
V_s=\argmin_{\mathbf{W}\in\mathcal{W},\bm{\Psi}\in\mathcal{M}_i}\,\,\, \mathrm{tr}(\mathbf{F}^{-1}). 
\end{equation}
It is obvious that problem \eqref{p:com-only} and \eqref{p:sen-only} are two special cases of \eqref{p:joint optimize} by setting $\rho = 1$ and $\rho = 0$, respectively. 
For brevity, the remainder of this work focuses solely on \eqref{p:joint optimize}, of which the optimization method can be readily applied to \eqref{p:com-only} and \eqref{p:sen-only}. 
One key challenge in solving problem \eqref{p:joint optimize} lies in the strong coupling between the scattering matrix $\bm{\Psi}$ and the active beamforming matrix $\mathbf{W}$ in both SINRs and the FIM. Additionally, the variables in the FIM exhibit a complicated blockwise structure, and the symmetric and unitary constraint imposed on $\bm{\Psi}$ is non-convex.
One common method to tackle these issues is to leverage classic iterative algorithms, which, however, inevitably lead to high computational complexity.
Hence, we propose an efficient AO algorithm that guarantees convergence to a locally optimal solution of \eqref{p:joint optimize} while maintaining very low computational complexity.    
\section{Proposed Joint Optimization Framework}\label{sec:3_algor}
This section develops an efficient AO method for problem \eqref{p:joint optimize}, where the scattering matrix $\bm{\Psi}$ and the active beamforming matrix $\mathbf{W}$ are iteratively optimized until convergence. 
Inspired by the SCA-based approach proposed in \cite{fty25_ISAClowcomplexity} for ISAC active beamforming design without RIS, we extend its application to subproblems of both $\bm{\Psi}$ and $\mathbf{W}$ in this considered BD-RIS-aided ISAC scenario. 
Unlike conventional iterative algorithms that solve each subproblem using interior-point techniques (typically implemented via optimization toolboxes such as CVX), the key novelty of the proposed PSCA approach lies in reformulating each subproblem into a series of more tractable problems with closed-form solutions. This significantly reduces computational complexity. Notably, the proposed approach can also be applied to conventional user-side diagonal RIS-aided ISAC resource allocation problems, offering much lower computational complexity.
The proposed AO-PSCA algorithm is elaborated in the following subsections.
\subsection{Scattering Matrix Optimization}\label{sec:3_A}
When $\mathbf{W}$ is fixed, the subproblem of \eqref{p:joint optimize} to optimize $\bm{\Psi}$ is formulated as
\begin{equation}\label{p:sub psi}
\mathop{\max}_{\bm{\Psi}\in\mathcal{M}_i}\,\,\, \rho\frac{R_{\mathrm{sum}}}{V_c}+(1-\rho)\frac{-\mathrm{tr}(\mathbf{F}^{-1})}{V_s}.
\end{equation}
We next propose a PSCA method to solve problem \eqref{p:sub psi}, which involves two steps at each iteration: \textit{i)} quadratic approximation: approximating the non-convex objective function with a tractable quadratic surrogate function, and \textit{ii)} closed-form projection: reformulating the problem into a projection form that enables a closed-form solution, thereby reducing the overall computational complexity. The two steps are detailed as follows:
\par
\subsubsection{Quadratic approximation}
We first utilize two lemmas to reformulate the non-convex $R_{\mathrm{sum}}$ and $-\mathrm{tr}(\mathbf{F}^{-1})$ in \eqref{p:sub psi} as more tractable quadratic forms.
\par
\begin{lemma}[\hspace{-0.2mm}\cite{fty23_Rk}]\label{lem:1}
    Let $s \in \mathbb{C}$ and $n \in \mathbb{R}_{++}$, for any given $(s_0,n_0)$ in the feasible domain, the function $\log\left(1+\frac{|s|^2}{n}\right)$ is lower bounded by its first-order Taylor expansion as
\begin{equation}
    \begin{aligned}
    \log\left(1+\frac{|s|^2}{n}\right)&\geq\log\left(1+\frac{|s_0|^2}{n_0} \right)-\frac{|s_0|^2}{n_0}+ 2\mathrm{Re}\left(\frac{s_0^\ast}{n_0}s\right)\\&-\frac{|s_0|^2}{n_0(n_0 +|s_0|^2)}(|s|^2+n),
    \end{aligned}
\end{equation}
where the equality is achieved when $(s,n)=(s_0,n_0)$.
\end{lemma}
\begin{lemma}[\hspace{-0.2mm}\cite{sy16_MM}]\label{lem:2}
    Let $\mathbf{V}\in\mathbb{S}_{++}$, for any given $\mathbf{V}_0$ in the feasible domain, the function $\mathrm{tr}(\mathbf{V}^{-1})$ is lower bounded by its first-order Taylor expansion as
\begin{equation}
    \mathrm{tr}(\mathbf{V}^{-1})\geq2\mathrm{tr}(\mathbf{V}^{-1}_0)-\mathrm{tr}(\mathbf{V}^{-1}_0\mathbf{V}^{-1}_0\mathbf{V}),
\end{equation}
where the equality is achieved when $\mathbf{V}=\mathbf{V}_0$.
\end{lemma}
\par
Following Lemma \ref{lem:1}, the SR expression $R_{\mathrm{sum}}=\sum_{k\in\mathcal{K}}\log(1+\gamma_k)$ with $\gamma_k$ specified in \eqref{eq:SINR} can be approximated at the $l$-th iteration as
\begin{equation}\label{eq:f_psi_R1}
\begin{aligned}
    &R_\mathrm{sum}\geq\sum_{k\in\mathcal{K}}\bigg[\log\Big(1+\gamma_k^{[l]}\Big)-\gamma_k^{[l]}+2\mathrm{Re}\Big(\zeta_k^{[l]}\mathbf{h}_{k}^H\bm{\Psi}\mathbf{H}\mathbf{w}_{k}\Big)\\
    &-\eta_k^{[l]}\Big(\sum_{i\in\mathcal{K}}|\mathbf{h}_{k}^H\bm{\Psi}\mathbf{H} \mathbf{w}_{i}|^2+\|\mathbf{h}_{k}^H\bm{\Psi}\mathbf{H}\mathbf{W}_s\|_2^2+\sigma_c^2\Big)\bigg]\triangleq{f_{C_\psi}^{[l]}},\\
\end{aligned}
\end{equation}
where $\zeta_{k}^{[l]}$ and $\eta_k^{[l]}, \forall k$ are auxiliary variables given by
\begin{equation}\label{eq:psi_r_zeta_eta}
    \begin{aligned}
        &\zeta_k^{[l]}\triangleq\frac{\gamma_k^{[l]}}{\mathbf{h}_{k}^H\bm{\Psi}^{[l]}\mathbf{H} \mathbf{w}_{k}}, \forall k,\\
        &\eta_k^{[l]}\triangleq\frac{\gamma_k^{[l]}}{\sum_{i\in\mathcal{K}}|\mathbf{h}_{k}^H\bm{\Psi}^{[l]}\mathbf{H} \mathbf{w}_{i}|^2+\|\mathbf{h}_{k}^H\bm{\Psi}^{[l]}\mathbf{H}\mathbf{W}_s\|_2^2+\sigma_c^2},\forall k.
    \end{aligned}
\end{equation}
By further defining the following matrices
\begin{equation}\label{eq:H_C_E}
\begin{aligned}
    &\mathbf{H}_c\triangleq[\mathbf{h}_1,\dots,\mathbf{h}_K],\ \ \mathbf{C}\triangleq\mathbf{H}\mathbf{W}\mathbf{W}^H\mathbf{H}^H,\\
    &\mathbf{E}_1^{[l]}\triangleq\mathrm{diag}(\zeta_1^{[l]},\dots,\zeta_K^{[l]}),\ \ \mathbf{E}_2^{[l]}\triangleq\mathrm{diag}(\eta_1^{[l]},\dots,\eta_K^{[l]}),
\end{aligned}
\end{equation}
the surrogate function of $R_\mathrm{sum}$ at the $l$-th iteration specified in \eqref{eq:f_psi_R1} can be equivalently transformed into
\begin{equation}\label{eq:f_psi_R2}
\begin{aligned}
    f_{C_\psi}^{[l]}&\triangleq 
    2\mathrm{Re}\left\{\mathrm{tr}\Big(\mathbf{E}_1^{[l]}\mathbf{H}_{c}^H\bm{\Psi}\mathbf{H}\mathbf{W}_c\Big)\right\}-\mathrm{tr}\Big(\mathbf{E}_2^{[l]}\mathbf{H}_{c}^H\bm{\Psi}\mathbf{C}\bm{\Psi}^H\mathbf{H}_c\Big)\\
    &+\sum_{k\in\mathcal{K}}\Big\{\log(1+\gamma_k^{[l]})-\gamma_k^{[l]}-\eta_k^{[l]}\sigma_c^2\Big\}.
\end{aligned}
\end{equation}
\par
\begin{figure*}[t]
    \centering
\begin{equation}\label{eq:Sigma}
\scalebox{0.92}{$
	\begin{aligned}
\bm{\Sigma}_{11}^{[l]}&\triangleq\mathbf{A}^\ast\mathbf{U}^H\big((\dot{\mathbf{B}}_{\bm{\theta}}^H \dot{\mathbf{B}}_{\bm{\theta}})\odot \mathbf{J}_{11}^{[l]}\big)\mathbf{U}\mathbf{A}^{T}+\mathbf{A}^\ast\mathbf{U}^H\big((\dot{\mathbf{B}}_{\bm{\theta}}^H\mathbf{B})\odot\mathbf{J}_{11}^{[l]} \big)\mathbf{U}\dot{\mathbf{A}}_{\bm{\theta}}^{T}+\dot{\mathbf{A}}_{\bm{\theta}}^\ast\mathbf{U}^H\big((\mathbf{B}^{H}\dot{\mathbf{B}}_{\bm{\theta}})\odot\mathbf{J}_{11}^{[l]} \big)\mathbf{U}\mathbf{A}^{T}+\dot{\mathbf{A}}_{\bm{\theta}}^\ast\mathbf{U}^H\big((\mathbf{B}^{H}\mathbf{B})\odot \mathbf{J}_{11}^{[l]}\big)\mathbf{U}\dot{\mathbf{A}}_{\bm{\theta}}^{T},\\
\bm{\Sigma}_{12}^{[l]}&\triangleq\mathbf{A}^\ast\mathbf{U}^H\big((\dot{\mathbf{B}}_{\bm{\theta}}^H \dot{\mathbf{B}}_{\bm{\phi}})\odot \mathbf{J}_{12}^{[l]}\big)\mathbf{U}\mathbf{A}^{T}+\mathbf{A}^\ast\mathbf{U}^H\big((\dot{\mathbf{B}}_{\bm{\theta}}^H\mathbf{B})\odot\mathbf{J}_{12}^{[l]} \big)\mathbf{U}\dot{\mathbf{A}}_{\bm{\phi}}^{T}+\dot{\mathbf{A}}_{\bm{\theta}}^\ast\mathbf{U}^H\big((\mathbf{B}^{H}\dot{\mathbf{B}}_{\bm{\phi}})\odot\mathbf{J}_{12}^{[l]} \big)\mathbf{U}\mathbf{A}^{T}+\dot{\mathbf{A}}_{\bm{\theta}}^\ast\mathbf{U}^H\big((\mathbf{B}^{H}\mathbf{B})\odot \mathbf{J}_{12}^{[l]}\big)\mathbf{U}\dot{\mathbf{A}}_{\bm{\phi}}^{T},\\
\bm{\Sigma}_{22}^{[l]}&\triangleq\mathbf{A}^\ast\mathbf{U}^H\big((\dot{\mathbf{B}}_{\bm{\phi}}^H \dot{\mathbf{B}}_{\bm{\phi}})\odot \mathbf{J}_{22}^{[l]}\big)\mathbf{U}\mathbf{A}^{T}+\mathbf{A}^\ast\mathbf{U}^H\big((\dot{\mathbf{B}}_{\bm{\phi}}^H\mathbf{B})\odot\mathbf{J}_{22}^{[l]} \big)\mathbf{U}\dot{\mathbf{A}}_{\bm{\phi}}^{T}+\dot{\mathbf{A}}_{\bm{\phi}}^\ast\mathbf{U}^H\big((\mathbf{B}^{H}\dot{\mathbf{B}}_{\bm{\phi}})\odot\mathbf{J}_{22}^{[l]} \big)\mathbf{U}\mathbf{A}^{T}+\dot{\mathbf{A}}_{\bm{\phi}}^\ast\mathbf{U}^H\big((\mathbf{B}^{H}\mathbf{B})\odot \mathbf{J}_{22}^{[l]}\big)\mathbf{U}\dot{\mathbf{A}}_{\bm{\phi}}^{T},\\
\bm{\Sigma}_{13}^{[l]}&\triangleq\mathbf{A}^\ast\mathbf{U}^H\big((\dot{\mathbf{B}}_{\bm{\theta}}^H \mathbf{B})\odot (\mathbf{J}_{13}^{[l]}+j\mathbf{J}_{14}^{[l]})\big)\mathbf{A}^{T}+\dot{\mathbf{A}}_{\bm{\theta}}^\ast\mathbf{U}^H\big((\mathbf{B}^H\mathbf{B})\odot(\mathbf{J}_{13}^{[l]}+j\mathbf{J}_{14}^{[l]}) \big)\mathbf{A}^{T},\\
\bm{\Sigma}_{23}^{[l]}&\triangleq\mathbf{A}^\ast\mathbf{U}^H\big((\dot{\mathbf{B}}_{\bm{\phi}}^H \mathbf{B})\odot (\mathbf{J}_{23}^{[l]}+j\mathbf{J}_{24}^{[l]})\big)\mathbf{A}^{T}+\dot{\mathbf{A}}_{\bm{\phi}}^\ast\mathbf{U}^H\big((\mathbf{B}^H\mathbf{B})\odot(\mathbf{J}_{23}^{[l]}+j\mathbf{J}_{24}^{[l]}) \big)\mathbf{A}^{T},\\
\bm{\Sigma}_{33}^{[l]}&\triangleq\mathbf{A}^\ast\big((\mathbf{B}^H \mathbf{B})\odot (\mathbf{J}_{33}^{[l]}+\mathbf{J}_{44}^{[l]}+2j\mathbf{J}_{34}^{[l]})\big)\mathbf{A}^{T}.
	\end{aligned}$}
\end{equation}
\rule{\linewidth}{0.5pt}
\end{figure*}
Also, by leveraging Lemma \ref{lem:2}, the non-convex sensing objective function $-\mathrm{tr}(\mathbf{F}^{-1})$ is approximated at the $l$-th iteration as 
\begin{equation}\label{eq:f_psi_S1}
    -\mathrm{tr}(\mathbf{F}^{-1})\leq-2\mathrm{tr}\{(\mathbf{F}^{[l]})^{-1}\}+\mathrm{tr}\{\mathbf{J}^{[l]}\mathbf{F}\}\triangleq{f_{S_\psi}^{[l]}},
\end{equation}
where we define $\mathbf{J}^{[l]}\triangleq(\mathbf{F}^{[l]})^{-2}$.
To deal with the complicated blockwise structure of $\bm{\Psi}$ in FIM $\mathbf{F}$, we further partition $\mathbf{J}^{[l]}$ into
\begin{equation}\label{eq:J}
	\mathbf{J}^{[l]}=\begin{bmatrix}
		\mathbf{J}_{11}^{[l]}&\mathbf{J}_{12}^{[l]}&\mathbf{J}_{13}^{[l]}&\mathbf{J}_{14}^{[l]}\\ 
		(\mathbf{J}_{12}^{[l]})^T&\mathbf{J}_{22}^{[l]}&\mathbf{J}_{23}^{[l]}&\mathbf{J}_{24}^{[l]}\\
		(\mathbf{J}_{13}^{[l]})^T&(\mathbf{J}_{23}^{[l]})^T&\mathbf{J}_{33}^{[l]}&\mathbf{J}_{34}^{[l]}\\
		(\mathbf{J}_{14}^{[l]})^T&(\mathbf{J}_{24}^{[l]})^T&(\mathbf{J}_{34}^{[l]})^T&\mathbf{J}_{44}^{[l]}
	\end{bmatrix}.
\end{equation}
This therefore facilitates a compact and tractable reformulation of the original surrogate function in \eqref{eq:f_psi_S1} at iteration $l$ as
\begin{equation}\label{eq:f_psi_S2}
    f_{S_\psi}^{[l]}\triangleq -2\mathrm{tr}\big\{\big(\mathbf{F}^{[l]}\big)^{-1}\big\}+\mathrm{Re}\big\{\mathrm{tr}\big(\bm{\Psi}\mathbf{C}\bm{\Psi}^H\mathbf{\Sigma}^{[l]}\big)\big\},
\end{equation}
where
\begin{equation}\label{eq:sigma_all}
    \bm{\Sigma}^{[l]}\triangleq\frac{2M}{\sigma_s^2}\left(\bm{\Sigma}_{11}^{[l]}+2\bm{\Sigma}_{12}^{[l]}+2\bm{\Sigma}_{13}^{[l]}+\bm{\Sigma}_{22}^{[l]}+2\bm{\Sigma}_{23}^{[l]}+\bm{\Sigma}_{33}^{[l]}\right),
\end{equation}
and $\bm{\Sigma}_{11}^{[l]}$, $\bm{\Sigma}_{12}^{[l]}$, $\bm{\Sigma}_{13}^{[l]}$, $\bm{\Sigma}_{22}^{[l]}$, $\bm{\Sigma}_{23}^{[l]}$, $\bm{\Sigma}_{33}^{[l]}$ are specified in \eqref{eq:Sigma} at the top of this page.
The detailed derivation of \eqref{eq:f_psi_S2} is specified in Appendix \ref{appendix B}.
\par
Using equations \eqref{eq:f_psi_R2} and \eqref{eq:f_psi_S2}, problem \eqref{p:sub psi} can be decomposed into a sequence of subproblems. At the $l$-th iteration, given $\bm{\Psi}^{[l]}$, the subproblem becomes
\begin{equation}\label{p:sub psi_sca1}
\mathop{\max}_{\bm{\Psi}\in\mathcal{M}_i}\,\,\, \frac{2\rho}{V_c}\mathrm{Re}\left\{\mathrm{tr}\Big(\mathbf{E}_1^{[l]}\mathbf{H}_{c}^H\bm{\Psi}\mathbf{H}\mathbf{W}_c\Big)\right\}+\mathrm{tr}\left(\mathbf{P}_1^{[l]}\bm{\Psi}\mathbf{C}\bm{\Psi}^H\right),
\end{equation}
where constant terms are omitted for brevity, and $\mathbf{P}_1^{[l]}$ is defined as
\begin{equation}\label{eq:P1}
    \mathbf{P}_1^{[l]}\triangleq\frac{-\rho}{V_c}\mathbf{H}_c\mathbf{E}_2^{[l]}\mathbf{H}_c^H + \frac{1-\rho}{2V_s}\left(\bm{\Sigma}^{[l]}+\bm{\Sigma}^{[l]^H}\right).
\end{equation}
As observed, the non-convexity of \eqref{p:sub psi_sca1} primarily stems from two factors: \textit{i)} the symmetric and unitary constraint $\bm{\Psi}\in\mathcal{M}_i$ is non-convex, \textit{ii)} the coefficient matrix with respect to $\bm{\Psi}\mathbf{C}\bm{\Psi}^H, \mathbf{C}\succeq \mathbf{0}_{N_I}$ in the objective function is not necessarily negative semidefinite, i.e., $\mathbf{P}_1^{[l]}\npreceq \mathbf{0}_{N_I}$.
Fortunately, although \eqref{p:sub psi_sca1} remains non-convex, its objective function exhibits a quadratic structure, which facilitates an efficient solution under the symmetric and unitary constraint, as detailed in the next step.
\begin{remark}
    It is worth noting that the combined surrogate function at the $l$-th iteration, i.e., $\frac{\rho}{V_c}f_{C_\psi}^{[l]}+\frac{1-\rho}{V_s}f_{S_\psi}^{[l]}$, does not serve as a lower bound of the original objective function, since $f_{C_\psi}^{[l]}$ specified in \eqref{eq:f_psi_R2} is the lower bound of $R_{\mathrm{sum}}$, while $f_{S_\psi}^{[l]}$ given by \eqref{eq:f_psi_S2} acts as the upper bound of $-\mathrm{tr}(\mathbf{F}^{-1})$. This therefore distinguishes our approach from standard SCA frameworks, which typically construct a lower-bound surrogate function to ensure convergence. Nevertheless, we will show in Section \ref{sec:3_D} that the proposed method still guarantees convergence.
\end{remark}
\subsubsection{Closed-form projection}
We first introduce a pre-defined constant term $\mu_1\mathbf{I}_{N_I}$ to transform its objective function into a convex form, i.e., $\mathbf{P}_1^{[l]}+\mu_1\mathbf{I}_{N_I}\succeq \mathbf{0}_{N_I}$. 
Given that $\bm{\Psi}\bm{\Psi}^H=\mathbf{I}_{N_I}$ holds with $\bm{\Psi}\in\mathcal{M}_i$, problem \eqref{p:sub psi_sca1} is equivalently reformulated as
\begin{equation}\label{p:sub psi_sca2}
\begin{aligned}
    \mathop{\max}_{\bm{\Psi}\in\mathcal{M}_i}\,\,\, 2\mathrm{Re}\left\{\mathrm{tr}\left(\bm{\Psi}^H\mathbf{P}_2^{[l]}\right)\right\}+\mathrm{tr}\left(\bm{\Psi}\mathbf{C}\bm{\Psi}^H(\mathbf{P}_1^{[l]}+\mu_1\mathbf{I}_{N_I})\right),
\end{aligned}
\end{equation}
where we define $\mathbf{P}_2^{[l]}\triangleq\frac{\rho}{V_c}\mathbf{H}_c(\mathbf{E}_1^{[l]})^H\mathbf{W}_c^H\mathbf{H}^H$. Due to the convexity of its objective function, problem \eqref{p:sub psi_sca2} is then efficiently solved by the following lemma.
\begin{lemma}[\hspace{-0.2mm}\cite{zxh25_jointpassive}]\label{lem:3}
    Let $\mathbf{Z},\mathbf{S}\in\mathbb{C}^{N_I\times N_I}$ be any given positive semidefinite Hermitian matrices, for any $\bm{\Psi}\in\mathbb{C}^{N_I\times N_I}$, we have
    \begin{equation}
        \mathrm{tr}(\bm{\Psi}\mathbf{Z}\bm{\Psi}^H\mathbf{S})\geq2\mathrm{Re}\{\mathrm{tr}(\bm{\Theta}\mathbf{Z}\bm{\Psi}^H\mathbf{S})\}-\mathrm{tr}(\bm{\Theta}\mathbf{Z}\bm{\Theta}^H\mathbf{S}),
    \end{equation}
    where $\bm{\Theta} \in \mathbb{C}^{N_I\times N_I}$ is an auxiliary matrix, and the equality is achieved at $\bm{\Theta}=\bm{\Psi}$. 
\end{lemma}
Based on Lemma \ref{lem:3}, we incorporate the auxiliary variable matrix $\bm{\Theta}$, which reformulates subproblem \eqref{p:sub psi_sca2} as \eqref{p:sub psi_sca2_} at the top of this page.
\begin{figure*}[t]
\centering
\begin{equation}\label{p:sub psi_sca2_}
\mathop{\max}_{\bm{\Psi},\bm{\Theta}\in\mathcal{M}_i}\,\,\, 2\mathrm{Re}\left\{\mathrm{tr}\big(\bm{\Psi}^H\mathbf{P}_2^{[l]}\big)\right\}+2\mathrm{Re}\Big\{\mathrm{tr}\big(\bm{\Theta}\mathbf{C}\bm{\Psi}^H(\mathbf{P}_1^{[l]}+\mu_1\mathbf{I}_{N_I})\big)\Big\}-\mathrm{tr}\left(\bm{\Theta}\mathbf{C}\bm{\Theta}^H(\mathbf{P}_1^{[l]}+\mu_1\mathbf{I}_{N_I})\right).
\end{equation}
\rule{\linewidth}{0.5pt}
\end{figure*}
Note that subproblem \eqref{p:sub psi_sca2_} exhibits a block-wise convex structure, where the objective function is convex with respect to either $\bm{\Theta}$ or $\bm{\Psi}$ when the other is fixed. This therefore enables $\bm{\Theta}$ and $\bm{\Psi}$ to be iteratively optimized with closed-form solutions.
To be specific, at the $l$-th iteration, we update $\bm{\Theta}^{[l]}=\bm{\Psi}^{[l]}$ according to the equality condition in Lemma \ref{lem:3}. With this given $\bm{\Theta}^{[l]}$, we then optimize $\bm{\Psi}^{[l+1]}$ via the approximated linear surrogate as
\begin{equation}\label{p:sub psi_sca3}
\mathop{\max}_{\bm{\Psi}\in\mathcal{M}_i}\,\,\, 2\mathrm{Re}\left[\mathrm{tr}\left\{\bm{\Psi}^H\Big(\mathbf{P}_2^{[l]}+\big(\mathbf{P}_1^{[l]}+\mu_1\mathbf{I}_{N_I}\big)\bm{\Theta}^{[l]}\mathbf{C}\Big)\right\}\right].
\end{equation}
To obtain the optimal solution of \eqref{p:sub psi_sca3}, we introduce the following definition and proposition.
\begin{definition}\label{def:1}
For an arbitrary square matrix $\mathbf{Q}\in\mathbb{C}^{N_I\times N_I}$, the projection operator that projects $\mathbf{Q}$ onto $\mathcal{M}_3$ is defined as
\begin{equation} \label{eq:proj_psi2}
    \bm{\Pi}_{\mathcal{M}_3}(\mathbf{Q})\triangleq \mathrm{diag}\left\{\mathrm{symuni}(\mathbf{Q}_1), \dots, \mathrm{symuni}(\mathbf{Q}_G)\right\},
\end{equation}
where $\mathbf{Q}_1, \dots,\mathbf{Q}_G$ are obtained as $\mathrm{blkdiag}\{\mathbf{Q}_1, \dots,\mathbf{Q}_G\}\triangleq\mathrm{blkdiag}\{\mathbf{1}_{N_{I,1}},\dots,\mathbf{1}_{N_{I,G}}\} \odot \mathbf{Q}$, and symuni is symmetric and unitary projection proposed in \cite{fty24_BDRISlowcomplexity}. It is defined as
\begin{equation}\label{eq:proj_psi1}
    \mathrm{symuni}(\mathbf{Q}_g)\triangleq\arg\min_{\mathbf{Z}^H\mathbf{Z}=\mathbf{I}, \mathbf{Z}=\mathbf{Z}^T}\|\mathbf{Q}_g-\mathbf{Z}\|_F^2 = \tilde{\mathbf{U}} \mathbf{V}^H,    
\end{equation}
where for $\mathbf{Q}_g\in\mathbb{C}^{N_{I,g}\times N_{I,g}}, \forall g$, we suppose that $\tilde{\mathbf{Q}}\triangleq\mathbf{Q}_g+\mathbf{Q}_g^T$ has a rank of $R$. Let its singular value decomposition (SVD) be $\tilde{\mathbf{Q}} = \mathbf{U}\mathbf{S}\mathbf{V}^H$, where $\mathbf{U}, \mathbf{V}$ can be partitioned as $\mathbf{U} = [\mathbf{U}_R, \mathbf{U}_{N_{I,g}-R}], \mathbf{V}=[\mathbf{V}_R, \mathbf{V}_{N_{I,g}-R}]$ accordingly. Then, we define $\tilde{\mathbf{U}}\triangleq [\mathbf{U}_R, \mathbf{V}^\ast_{N_{I,g}-R}]$. 
Note that when $G = N$, the operator $\bm{\Pi}_{\mathcal{M}_3}$ reduces to the projection onto $\mathcal{M}_1$, i.e., $\bm{\Pi}_{\mathcal{M}_1}$, whereas when $G = 1$, it becomes $\bm{\Pi}_{\mathcal{M}_2}$. 
\end{definition}
\begin{proposition}\label{prop:1}
    The optimal solution to \eqref{p:sub psi_sca3} is given by
    \begin{equation}\label{eq:optimal psi}
        \bm{\Psi}^\star=\bm{\Pi}_{\mathcal{M}_i}\Big(\mathbf{P}_2^{[l]}+\big(\mathbf{P}_1^{[l]}+\mu_1\mathbf{I}_{N_I}\big)\bm{\Theta}^{[l]}\mathbf{C}\Big).
    \end{equation}
\end{proposition}
\textit{Proof:} Since $\bm{\Psi}\bm{\Psi}^H=\mathbf{I}_{N_I}$ holds with $\bm{\Psi}\in\mathcal{M}_i$, problem \eqref{p:sub psi_sca3} is equivalent to the following projection problem
\begin{equation}\label{p:sub psi_sca4}
\mathop{\min}_{\bm{\Psi}\in\mathcal{M}_i}\,\,\, \Big\|\bm{\Psi}-\Big(\mathbf{P}_2^{[l]}+\big(\mathbf{P}_1^{[l]}+\mu_1\mathbf{I}_{N_I}\big)\bm{\Theta}^{[l]}\mathbf{C}\Big)\Big\|_F^2.
\end{equation}
Based on \eqref{eq:proj_psi2} and \eqref{eq:proj_psi1}, the optimal solution $\bm{\Psi}$ to \eqref{p:sub psi_sca4} is easily obtained as \eqref{eq:optimal psi}, which completes the proof.\hfill $\blacksquare$
\par
Based on the above derivation, the proposed PSCA approach for solving problem \eqref{p:sub psi} is summarized below, with the iteration index $l$ omitted for simplicity:
\begin{itemize}
    \item[1)] Initialize $\bm{\Psi}\in\mathcal{M}_i$;
    \item[2)] Update auxiliary variables $\gamma_k, \zeta_k, \eta_k, \forall k$ and $\mathbf{J}=\mathbf{F}^{-2}$ by \eqref{eq:SINR}, \eqref{eq:psi_r_zeta_eta} and \eqref{eq:FIM};
    \item[3)] Update $\mathbf{\bm{\Theta}}=\bm{\Psi}$;
    \item[4)] Update $\bm{\Psi}=\bm{\Pi}_{\mathcal{M}_i}\Big(\mathbf{P}_2+\big(\mathbf{P}_1+\mu_1\mathbf{I}_{N_I}\big)\bm{\Theta}\mathbf{C}\Big)$;
    \item[5)] Return to 2) until convergence.
\end{itemize}
\subsection{Active Beamforming Optimization}\label{sec:3_B}
With a given $\bm{\Psi}$, the subproblem of \eqref{p:joint optimize} with respect to $\mathbf{W}$ is expressed as
\begin{equation}\label{p:sub w}
\mathop{\max}_{\mathbf{W}\in\mathcal{W}}\,\,\, \rho\frac{R_{\mathrm{sum}}}{V_c}+(1-\rho)\frac{-\mathrm{tr}(\mathbf{F}^{-1})}{V_s}.
\end{equation}
Since this objective function with respect to $\mathbf{W}$ shares the same mathematical structure as the one in \eqref{p:sub psi}, the same PSCA method is applied to solve \eqref{p:sub w} as follows.
\par
\subsubsection{Quadratic approximation} 
Again, we first apply Lemma \ref{lem:1} and \ref{lem:2} to transform the non-convex $R_{\mathrm{sum}}$ and $-\mathrm{tr}(\mathbf{F}^{-1})$ into more tractable and explicit quadratic forms.
To be specific, according to Lemma \ref{lem:1}, $R_{\mathrm{sum}}=\sum_{k\in\mathcal{K}}\log(1+\gamma_k)$ is lower bounded by
\begin{equation}\label{eq:f_R2}
\begin{aligned}
    f_{C_w}^{[t]}&\triangleq 
    2\mathrm{Re}\left\{\mathrm{tr}\Big(\tilde{\mathbf{E}}_1^{[t]}\mathbf{G}^H\mathbf{W}_c\Big)\right\}-\mathrm{tr}\Big(\tilde{\mathbf{E}}_2^{[t]}\mathbf{G}^H\mathbf{W}\mathbf{W}^H\mathbf{G}\Big)\\
    &+\sum_{k\in\mathcal{K}}\Big\{\log(1+\tilde{\gamma}_k^{[t]})-\tilde{\gamma}_k^{[t]}-\tilde{\eta}_k^{[t]}\sigma_c^2\Big\},
\end{aligned}
\end{equation}
where $t$ denotes the $t$-th iteration of this subproblem with respect to $\mathbf{W}$, and $\mathbf{G}\triangleq[\mathbf{g}_1,\dots,\mathbf{g}_K]$ is the compact effective channel matrix. The auxiliary variables $\tilde{\gamma}_k^{[t]}, \tilde{\zeta}_k^{[t]}, \tilde{\eta}_k^{[t]}, \forall k$, and $\tilde{\mathbf{E}}^{[t]}_1, \tilde{\mathbf{E}}^{[t]}_2$ are given by
\begin{equation}\label{eq:r_zeta_eta}
    \begin{aligned}
        &\tilde{\gamma}_k^{[t]}\triangleq\frac{|\mathbf{g}_{k}^H\mathbf{w}^{[t]}_{k}|^2}{\sum_{i\neq k}^K|\mathbf{g}_{k}^H \mathbf{w}^{[t]}_{i}|^2+\|\mathbf{g}_{k}^H\mathbf{W}^{[t]}_s\|_2^2+\sigma_c^2}, \forall k,\\
        &\tilde{\zeta}_k^{[t]}\triangleq\frac{\tilde{\gamma}_k^{[t]}}{\mathbf{g}_{k}^H \mathbf{w}_{k}^{[t]}}, \forall k,\ \ \  \tilde{\eta}_k^{[t]}\triangleq\frac{\tilde{\gamma}_k^{[t]}}{\|\mathbf{g}_{k}^H\mathbf{W}^{[t]}\|_2^2+\sigma_c^2},\forall k,\\
        &\tilde{\mathbf{E}}_1^{[t]}\triangleq\mathrm{diag}(\tilde{\zeta}_1^{[t]},\dots,\tilde{\zeta}_K^{[t]}),\ \ \tilde{\mathbf{E}}_2^{[t]}\triangleq\mathrm{diag}(\tilde{\eta}_1^{[t]},\dots,\tilde{\eta}_K^{[t]}).
    \end{aligned}
\end{equation}
\par
Furthermore, based on Lemma \ref{lem:2} and the blockwise partition of $\tilde{\mathbf{J}}^{[t]}\triangleq(\tilde{\mathbf{F}}^{[t]})^{-2}$, the sensing objective function $-\mathrm{tr}(\mathbf{F}^{-1})$ is reformulated at the $t$-th iteration as 
\begin{equation}\label{eq:f_S2}
    f_{S_w}^{[t]}\triangleq -2\mathrm{tr}\big\{\big(\tilde{\mathbf{F}}^{[t]}\big)^{-1}\big\}+\mathrm{Re}\big\{\mathrm{tr}\big(\mathbf{W}\mathbf{W}^H\mathbf{H}^H\bm{\Psi}^H\tilde{\mathbf{\Sigma}}^{[t]}\bm{\Psi}\mathbf{H}\big)\big\},
\end{equation}
where $\tilde{\mathbf{F}}$ and $\tilde{\bm{\Sigma}}$ retain the same forms as \eqref{eq:FIM} and \eqref{eq:sigma_all}. The only difference is that $\tilde{\mathbf{F}}^{[t]}$ and $\tilde{\bm{\Sigma}}^{[t]}$ are calculated based on a fixed $\bm{\Psi}$ and a given $\mathbf{W}^{[t]}$ at the $t$-th iteration.
\par
By leveraging the surrogate functions given by \eqref{eq:f_R2} and \eqref{eq:f_S2}, problem \eqref{p:sub w} with a given $\mathbf{W}^{[t]}$ at the $t$-th PSCA iteration becomes
\begin{equation}\label{p:sub w_sca1}
\mathop{\max}_{\mathbf{W}\in\mathcal{W}}\,\,\, \frac{2\rho}{V_c}\mathrm{Re}\left\{\mathrm{tr}\Big(\tilde{\mathbf{E}}_1^{[t]}\mathbf{G}^H\mathbf{W}_c\Big)\right\}+\mathrm{tr}\Big(\tilde{\mathbf{P}}_1^{[t]}\mathbf{W}\mathbf{W}^H\Big),
\end{equation}
where we omit constant terms for clarity, and $\tilde{\mathbf{P}}_1^{[t]}$ is given by
\begin{equation}
    \tilde{\mathbf{P}}_1^{[t]}\triangleq\frac{-\rho}{V_c}\mathbf{G}\tilde{\mathbf{E}}_2^{[t]}\mathbf{G}^H + \frac{1-\rho}{2V_s}\mathbf{H}^H\mathbf{\Psi}^H(\tilde{\bm{\Sigma}}^{[t]}+\tilde{\bm{\Sigma}}^{[t]^H})\bm{\Psi}\mathbf{H}.
\end{equation}
Notice that the non-convexity of problem \eqref{p:sub w_sca1} arises from the fact that the power constraint is non-convex, and the coefficient matrix of the quadratic objective is not necessarily negative semidefinite, i.e., $\tilde{\mathbf{P}}_1^{[t]}\npreceq \mathbf{0}_{N_T}$.
Notably, despite its non-convexity, problem \eqref{p:sub w_sca1} exhibits the structure of a quadratically constrained quadratic problem (QCQP), which can be efficiently solved as follows.
\par
\subsubsection{Closed-form projection}
Similarly, we first lift the coefficient matrix $\tilde{\mathbf{P}}_1^{[t]}$ to be positive semidefinite by leveraging a constant term $\mu_2\mathbf{I}_{N_T}$, i.e., $\tilde{\mathbf{P}}_1^{[t]}+\mu_2\mathbf{I}_{N_T}\succeq\mathbf{0}_{N_T}$, thus ensuring the convexity of the objective function. 
Since $\mathrm{tr}(\mathbf{W}\mathbf{W}^H)=P$ holds at the optimal solution, problem \eqref{p:sub w_sca1} is equivalently rewritten as
\begin{equation}\label{p:sub w_sca2}
\mathop{\max}_{\mathbf{W}\in\mathcal{W}}\,\,\, 2\mathrm{Re}\left\{\mathrm{tr}\left(\mathbf{W}^H\tilde{\mathbf{P}}_2^{[t]}\right)\right\}+\mathrm{tr}\left(\mathbf{W}\mathbf{W}^H(\tilde{\mathbf{P}}_1^{[t]}+\mu_2\mathbf{I}_{N_T})\right),
\end{equation}
where $\tilde{\mathbf{P}}_2^{[t]}\triangleq[\frac{\rho}{V_c}\mathbf{G}(\tilde{\mathbf{E}}_1^{[t]})^H,\mathbf{0}_{N_T\times N_T}]$. 
By further introducing the auxiliary variable matrix $\tilde{\mathbf{Q}}\in\mathbb{C}^{N_T\times (K+N_T)}$, we transform the $t$-th PSCA subproblem \eqref{p:sub w_sca2} into \eqref{p:sub w_sca2_} at the top of this page.
\begin{figure*}[t]
    \centering
\begin{equation}\label{p:sub w_sca2_}
\mathop{\max}_{\mathbf{W},\mathbf{\tilde{\mathbf{Q}}}\in\mathcal{W}}\,\,\, 2\mathrm{Re}\left\{\mathrm{tr}\big(\mathbf{W}^H\tilde{\mathbf{P}}_2^{[t]}\big)\right\}+2\mathrm{Re}\Big\{\mathrm{tr}\big(\tilde{\mathbf{Q}}\mathbf{W}^H(\tilde{\mathbf{P}}_1^{[t]}+\mu_2\mathbf{I}_{N_T})\big)\Big\}-\mathrm{tr}\big(\tilde{\mathbf{Q}}\tilde{\mathbf{Q}}^H(\tilde{\mathbf{P}}_1^{[t]}+\mu_2\mathbf{I}_{N_T})\big).
\end{equation}
\rule{\linewidth}{0.5pt}
\end{figure*}
In \eqref{p:sub w_sca2_}, $\tilde{\mathbf{Q}}$ and $\mathbf{W}$ are iteratively optimized at each iteration.
Based on Lemma \ref{lem:3}, we update $\tilde{\mathbf{Q}}^{[t]}=\mathbf{W}^{[t]}$ at the $t$-th iteration. Similar to Proposition \ref{prop:1}, with this given $\tilde{\mathbf{Q}}^{[t]}$, we optimize $\mathbf{W}^{[t+1]}$ as
\begin{equation}\label{eq:optimal w}
\mathbf{W}^{[t+1]}=\bm{\Pi}_{\mathcal{W}}\Big(\tilde{\mathbf{P}}_2^{[t]}+\big(\tilde{\mathbf{P}}_1^{[t]}+\mu_2\mathbf{I}_{N_T}\big)\tilde{\mathbf{Q}}^{[t]}\Big),
\end{equation}
where $\bm{\Pi}_{\mathcal{W}}$ is defined as the projection operator onto the set $\mathcal{W}$, i.e.,
\begin{equation}
    \bm{\Pi}_{\mathcal{W}}(\mathbf{Z})\triangleq\arg\min_{\mathbf{S}\in\mathcal{W}}\|\mathbf{Z}-\mathbf{S}\|_F^2=\sqrt{P/\mathrm{tr}(\mathbf{Z}\mathbf{Z}^H)}\mathbf{Z}.
\end{equation}
Note that the overall PSCA approach for solving \eqref{p:sub w} follows a procedure similar to that for problem \eqref{p:sub psi}, and is therefore omitted here for brevity.
\subsection[Overall Alternating Optimization Framework for joint optimize]{Overall Alternating Optimization Framework for \eqref{p:joint optimize}}
\begingroup
\setlength{\textfloatsep}{3pt}
\begin{algorithm}[t!]
    \textbf{Input}: Transmit power budget $P$, the channel matrices $\mathbf{H}$, $\mathbf{h}_k,\forall k$, and convergence tolerance $\tau$ \;
    \textbf{Initialize}: $p\leftarrow0$, $\bm{\Psi}^{[p]}$, $\mathbf{W}^{[p]}$, $\mathrm{Obj}^{[p]}$ \;
    \Repeat{$|\mathrm{Obj}^{[p]} - \mathrm{Obj}^{[p-1]}| \leq \tau$}{
    $p \leftarrow p + 1$\;
    \textbf{Initialize:} $l \leftarrow 0$, $\bm{\Psi}^{[l]} \leftarrow \bm{\Psi}^{[p-1]}$\;
    \Repeat{convergence}{
        Update $\gamma_k^{[l]}, \zeta_k^{[l]}, \eta_k^{[l]}, \forall k$ and $\mathbf{J}^{[l]}=(\mathbf{F}^{[l]})^{-2}$ by \eqref{eq:SINR}, \eqref{eq:psi_r_zeta_eta} and \eqref{eq:FIM}\;
        Update $\bm{\Theta}^{[l]}=\bm{\Psi}^{[l]}$\;
        Update $\bm{\Psi}^{[l+1]}$ by \eqref{eq:optimal psi}\;
        $l \leftarrow l + 1$\;
    }
    \textbf{Initialize:} $t \leftarrow 0$, $\mathbf{W}^{[t]} \leftarrow \mathbf{W}^{[p-1]}$\;
    \Repeat{convergence}{
        Update $\tilde{\gamma}_k^{[t]}, \tilde{\zeta}_k^{[t]}, \tilde{\eta}_k^{[t]}, \forall k$ and $\tilde{\mathbf{J}}^{[t]}=(\tilde{\mathbf{F}}^{[t]})^{-2}$ by \eqref{eq:r_zeta_eta} and \eqref{eq:FIM}\;
        Update $\tilde{\mathbf{Q}}^{[t]}=\mathbf{W}^{[t]}$\;
        Update $\mathbf{W}^{[t+1]}$ by \eqref{eq:optimal w}\;
        $t \leftarrow t + 1$\;
    }
    Update $\bm{\Psi}^{[p]}=\bm{\Psi}^{[l]}$, $\mathbf{W}^{[p]}=\mathbf{W}^{[t]}$, $\mathrm{Obj}^{[p]}$ \;
}
	\caption{The proposed AO-PSCA algorithm to solve \eqref{p:joint optimize}}
	\label{a:AO-SCA}
\end{algorithm}
Based on the iterative closed-form updates of $\bm{\Psi}$ and $\mathbf{W}$ derived via PSCA in Section \ref{sec:3_A} and \ref{sec:3_B}, we develop an efficient AO method for solving problem \eqref{p:joint optimize}. 
The overall AO-PSCA optimization procedure is detailed in Algorithm \ref{a:AO-SCA}.
With an initialized non-zero feasible scattering matrix $\bm{\Psi}^{[0]}$ and active beamforming matrix $\mathbf{W}^{[0]}$, we iteratively update $\bm{\Psi}^{[p]}$ and $\mathbf{W}^{[p]}$ via the PSCA method. This process continues until the objective value in \eqref{p:joint optimize} converges.
\subsection{Convergence Analysis}\label{sec:3_D}
This subsection establishes the convergence of the proposed AO algorithm, where each subproblem is solved via a PSCA method.
We show that for both subproblems of $\bm{\Psi}$ and $\mathbf{W}$, the corresponding objective value increases monotonically after each iteration, thereby guaranteeing the convergence.
The core idea of this analysis is that the proposed PSCA method essentially constructs a sequence of first-order linear lower bound maximization subproblems, each followed by a projection onto the feasible set.
To be specific, we denote the objective function of the original subproblem \eqref{p:sub psi} with respect to $\bm{\Psi}$ as $g(\bm{\Psi})\triangleq g_c(\bm{\Psi})+g_s(\bm{\Psi})$, and the objective function of problem \eqref{p:sub psi_sca3} as $h^{[l]}(\bm{\Psi})\triangleq h^{[l]}_c(\bm{\Psi})+h^{[l]}_s(\bm{\Psi})+2\mu_1\mathrm{Re}\left[\mathrm{tr}\left\{\bm{\Psi}^H\bm{\Psi}^{[l]}\mathbf{C}\right\}\right]$ with 
\vspace{-2mm}
\begin{equation}\label{eq:g_h}
    \begin{aligned}
        &g_c(\bm{\Psi})\triangleq \frac{\rho}{V_c}R_{\mathrm{sum}},\ \ g_s(\bm{\Psi})\triangleq \frac{-(1-\rho)}{V_s}\mathrm{tr}(\mathbf{F}^{-1}),\\
        &h^{[l]}_c(\bm{\Psi})\triangleq2\mathrm{Re}\left[\mathrm{tr}\left\{\bm{\Psi}^H\Big(\mathbf{P}_2^{[l]}+\frac{-\rho}{V_c}\mathbf{H}_c\mathbf{E}_2^{[l]}\mathbf{H}_c^H\bm{\Psi}^{[l]}\mathbf{C}\Big)\right\}\right],\\
        &h^{[l]}_s(\bm{\Psi})\triangleq\frac{1-\rho}{V_s}\mathrm{Re}\left[\mathrm{tr}\left(\bm{\Psi}^H(\bm{\Sigma}^{[l]}+\bm{\Sigma}^{[l]^H})\bm{\Psi}^{[l]}\mathbf{C}\right)\right],
    \end{aligned}
\end{equation}
\endgroup
where $\bm{\Theta}^{[l]}$ is substituted with $\bm{\Psi}^{[l]}$ according to Lemma \ref{lem:3}.
Note that, unless otherwise specified, in the subproblem of $\bm{\Psi}$ at the $(p+1)$-th AO iteration, the given $\mathbf{W}^{[p]}$ in each function notation (e.g., in the original objective $g(\bm{\Psi}, \mathbf{W}^{[p]})$) is omitted for simplicity. 
To proceed this analysis, we invoke the following proposition.
\begin{proposition}\label{prop:2}
    The gradient of $g(\bm{\Psi})$ at the point $\mathbf{\Psi}=\mathbf{\Psi}^{[l]}$ is calculated as
    \begin{equation}\label{eq:grad}
    \begin{aligned}
        \nabla g(\bm{\Psi})\big|_{\bm{\Psi}=\bm{\Psi}^{[l]}}&=\nabla g_c(\bm{\Psi})\big|_{\bm{\Psi}=\bm{\Psi}^{[l]}}+\nabla g_s(\bm{\Psi})\big|_{\bm{\Psi}=\bm{\Psi}^{[l]}}\\
        &=\nabla h_c^{[l]}(\bm{\Psi})\big|_{\bm{\Psi}=\bm{\Psi}^{[l]}}+\nabla h_s^{[l]}(\bm{\Psi})\big|_{\bm{\Psi}=\bm{\Psi}^{[l]}}\\
        &=2\mathbf{P}_2^{[l]}+2\mathbf{P}_1^{[l]}\bm{\Psi}^{[l]}\mathbf{C}.
    \end{aligned}
    \end{equation}
\end{proposition}
\textit{Proof:} Please refer to Appendix \ref{appendix C}. \hfill $\blacksquare$
\par
By leveraging Proposition \ref{prop:2} and the property $\bm{\Pi}_{\mathcal{M}_i}(\mu\mathbf{X})=\bm{\Pi}_{\mathcal{M}_i}(\mathbf{X})$ derived from $\mathrm{symuni}(\mu\tilde{\mathbf{X}})=\mathrm{symuni}(\tilde{\mathbf{X}})$ based on \eqref{eq:proj_psi1}, we rewrite \eqref{eq:optimal psi} as
\begin{equation}\label{eq: iterative psi}
    \begin{aligned}
        \bm{\Psi}^{[t+1]}&=\bm{\Pi}_{\mathcal{M}_i}\Big(\mathbf{P}_2^{[l]}+\mathbf{P}_1^{[l]}\bm{\Psi}^{[l]}\mathbf{C}+\mu_1\bm{\Psi}^{[l]}\mathbf{C}\Big)\\
        &=\bm{\Pi}_{\mathcal{M}_i}\Big(\nabla g(\bm{\Psi})\big|_{\bm{\Psi}=\bm{\Psi}^{[l]}}+2\mu_1\bm{\Psi}^{[l]}\mathbf{C}\Big), 
    \end{aligned}
\end{equation}
where the construction of surrogate functions in the PSCA method aligns with computing the gradient of both the original objective function $g(\bm{\Psi})$ and the term $\mu_1 \mathrm{tr}(\bm{\Psi} \mathbf{C} \bm{\Psi}^H)$ introduced in \eqref{p:sub psi_sca2}. 
This observation allows us to establish the convergence of the proposed method as follows.
\par
Given that $\bm{\Psi}\bm{\Psi}^H = \mathbf{I}_{N_I}$, and $\mathbf{C}$ is constant in subproblem \eqref{p:sub psi}, the corresponding subproblem can be equivalently reformulated as
\begin{equation}\label{p:sub psi_equal}
\mathop{\max}_{\bm{\Psi}\in\mathcal{M}_i}\,\,\, g(\bm{\Psi})+\beta\mathrm{tr}(\bm{\Psi} \mathbf{C} \bm{\Psi}^H).
\end{equation}
Since $\mathbf{C}\succeq\mathbf{0}_{N_I}$, $\beta$ is appropriately chosen to ensure the convexity of the reformulated objective function $c(\bm{\Psi})\triangleq g(\bm{\Psi})+\beta\mathrm{tr}(\bm{\Psi} \mathbf{C} \bm{\Psi}^H)$. 
\par
For the newly established convex $c(\bm{\Psi})$, we construct the linear lower bound via its first-order Taylor expansion at the 
$l$-th iteration, which is given by
\begin{equation}\label{eq:c(psi)bound}
\begin{aligned}
    c(\bm{\Psi})&\geq c(\bm{\Psi}^{[l]})+\langle\nabla c(\bm{\Psi})\big|_{\bm{\Psi}=\bm{\Psi}^{[l]}}, \bm{\Psi}-\bm{\Psi}^{[l]}\rangle\\
    &=\langle\nabla g(\bm{\Psi})\big|_{\bm{\Psi}=\bm{\Psi}^{[l]}}+2\beta\bm{\Psi}^{[l]}\mathbf{C}, \bm{\Psi}\rangle+c_\mathrm{const},
\end{aligned}
\end{equation}
where $c_\mathrm{const}$ denotes the remaining constant term. 
\par
We then maximize this linear lower bound, followed by a projection onto the boundary of $\mathcal{M}_i$, which yields the following iterative update:
\begin{equation}
    \bm{\Psi}^{[l+1]}=\bm{\Pi}_{\mathcal{M}_i}\Big(\nabla g(\bm{\Psi})\big|_{\bm{\Psi}=\bm{\Psi}^{[l]}}+2\beta\bm{\Psi}^{[l]}\mathbf{C}\Big).
\end{equation}
Observe that this coincides with the update rule given by \eqref{eq: iterative psi} when $\mu_1=\beta$. Hence, with $\mu_1\geq\beta$, we establish the final convergence analysis as
\begin{equation}
\begin{aligned}
    g(\bm{\Psi}^{[l+1]})&= c(\bm{\Psi}^{[l+1]})-\mu_1\mathrm{tr}(\bm{\Psi}^{[l+1]}\mathbf{C}(\bm{\Psi}^{[l+1]})^H)\\
    &\overset{(\mathrm{a})}{\geq} g(\bm{\Psi}^{[l]})+\langle\nabla c(\bm{\Psi})\big|_{\bm{\Psi}=\bm{\Psi}^{[l]}}, \bm{\Psi}^{[l+1]}-\bm{\Psi}^{[l]}\rangle\\
    &\overset{(\mathrm{b})}{\geq} g(\bm{\Psi}^{[l]}),
\end{aligned}
\end{equation}
where (a) comes from the definition of $c(\bm{\Psi})$ and the linear lower bound given by \eqref{eq:c(psi)bound}, (b) is obtained since $\langle \nabla 
c(\bm{\Psi})\big|_{\bm{\Psi}=\bm{\Psi}^{[l]}}, \bm{\Psi}^{[l+1]}-\bm{\Psi}^{[l]}\rangle \geq 0$ holds for the convex function $c(\bm{\Psi})$.
Given that the original objective function $g(\bm{\Psi})$ increases monotonically and $\bm{\Psi}$ is bounded over the closed set $\mathcal{M}_i$, we then guarantee the convergence of the subproblem with respect to $\bm{\Psi}$, i.e., $g(\bm{\Psi}^{[p+1]},\mathbf{W}^{[p]})\geq g(\bm{\Psi}^{[p]},\mathbf{W}^{[p]})$.
A similar convergence analysis holds for the subproblem of $\mathbf{W}$, leading to $g(\bm{\Psi}^{[p+1]},\mathbf{W}^{[p+1]})\geq g(\bm{\Psi}^{[p+1]},\mathbf{W}^{[p]})$.
This implies that the AO-PSCA algorithm maintains a monotonic increase of the objective as $g(\bm{\Psi}^{[p+1]}, \mathbf{W}^{[p+1]}) \ge g(\bm{\Psi}^{[p+1]}, \mathbf{W}^{[p]}) \ge g(\bm{\Psi}^{[p]}, \mathbf{W}^{[p]})$, thereby guaranteeing the convergence of the overall algorithm within a specified tolerance $\tau$.
\subsection{Computational Complexity Analysis}\label{sec:3_E}
The complexity of the AO method is determined by the PSCA approach for addressing the subproblems of $\bm{\Psi}$ and $\mathbf{W}$, which mainly involve matrix multiplications and inversions. 
Specifically, at each iteration of the PSCA algorithm, the update of auxiliary variables $\gamma_k, \zeta_k, \eta_k, \forall k$ incurs a complexity of $\mathcal{O}\big(K(N_T+K)(N_I^2+N_IN_T)\big)$. 
The complexity of computing the auxiliary variable $\mathbf{J}$ is $\mathcal{O}\big(N_T^2(K+N_T)+Q(N_I^2+N_TN_I+N_T^2)+Q^2(N_I+N_S)+Q^3\big)$. 
The update of the fully-connected BD-RIS scattering matrix involves a complexity of $\mathcal{O}(N_I^3+N_I^2(K+N_T)+QN_I^2+Q^2(N_I+N_S)+N_I(K^2+KN_T+N_T^2)$. 
Accordingly, the complexity for the subproblem with respect to $\bm{\Psi}$ is of order $\mathcal{O}\Big(I_\psi\big(N_I^3+Q(N_I^2+N_TN_I+N_T^2)+Q^2(N_I+N_S)+N_T^3+Q^3+K(N_T+K)(N_I^2+N_IN_T)\big)\Big)$, where $I_\psi$ denotes the maximum iteration numbers of all AO steps for optimizing $\bm{\Psi}$. 
Similarly, we obtain the complexity over $I_w$ iterations for updating $\mathbf{W}$ as $\mathcal{O}\Big(I_w\big(Q(N_I^2+N_TN_I+N_T^2)+Q^2(N_I+N_S)+N_T^3+Q^3+K(N_T+K)(N_I^2+N_IN_T)\big)\Big)$.
Hence, assuming the AO algorithm converges within $I_0$ iterations, the overall complexity is 
$\mathcal{O}\Big(I_0\big(I_\psi+I_w\big)\big(Q(N_I^2+N_TN_I+N_T^2)+Q^2(N_I+N_S)+N_T^3+Q^3+K(N_T+K)(N_I^2+N_IN_T)\big)+I_0I_\Psi N_I^3\Big)$.
This substantially lowers the dominant polynomial order by leveraging iterative closed-form updates, in sharp contrast to conventional methods such as WMMSE-PDD \cite{yzh_wmmsepdd}, which exhibits a higher per-iteration complexity of approximately $\mathcal{O}\big(N_I^2(N_I+K+N_T)^{4.5}\log(1/\tau)\big)$.
\section{Numerical Results}\label{sec:4_result}
In this section, we evaluate the performance of the transmitter-side BD-RIS-enabled ISAC system model as well as the proposed AO-PSCA algorithm.
Note that when scenarios involve different numbers of targets, we choose the average trace of CRB given by $\mathrm{tr}(\mathbf{F}^{-1})/Q$ rather than the trace of CRB to ensure a fair comparison. 
In this work, we adopt a 3-dimensional Cartesian coordinate system. 
Unless otherwise specified, the BD-RIS with $N_I=32$ elements is configured as a uniform planar array (UPA) lying on the $yz$ plane, while the $N_T = 4$ active antennas are arranged as a uniform linear array (ULA) parallel to the $y$ axis at the vertical center of the BD-RIS 10 wavelengths away. 
Moreover, the sensor with $N_S=6$ elements is positioned alongside the $y$ axis at the BD-RIS \cite{wzl23_STARS}.
A half-wavelength spacing is considered for adjacent active antennas, BD-RIS and sensor elements at a carrier frequency of $f_c = 30\ \mathrm{GHz}$.
This BD-RIS-aided ISAC system simultaneously serves $K=4$ communication users and senses $Q=2$ point targets over a CPI of length $M = 128$.
Unless otherwise specified, the transmit power budget is $P=6\ \mathrm{dBm}$, while the communication and sensing noise variances are $\sigma_c^2=\sigma_s^2=0\ \mathrm{dBm}$. The weighting factor is set to $\rho=0.8$.
The communication channel $\mathbf{h}_k, \forall k$ is assumed to follow a circularly symmetric complex Gaussian (CSCG) distribution.
In addition, we assume that three distinct targets are located at $[\theta_q, \phi_q]=[-30^{\circ}, 15^{\circ}]$,  $[10^{\circ}, -45^{\circ}]$ and $[-60^{\circ}, -75^{\circ}]$ with respect to the BD-RIS, respectively. 
The complex reflection coefficient $\alpha_q, \forall q$ is set as $(1+0.2n)e^{j2\pi n}$ with $n$ following a uniform distribution $\mathcal{U}(0,1)$ \cite{fty25_ISAClowcomplexity}.  
The power efficiency of BD-RIS $\rho_{\mathrm{eff}}$ in \eqref{eq:HT} is set to $\rho_{\mathrm{eff}}=1$, and the active and passive antenna gains involved in $\mathbf{H}$ are $G^A=G^B=3\ \mathrm{dB}$ \cite{Mishra23_BDRIScomm, buzzi21_transmitRIS}.
\par
\begingroup
\setlength{\textfloatsep}{0.2cm}
\begin{figure}[t]
    \centering
    \includegraphics[width=0.9\linewidth]{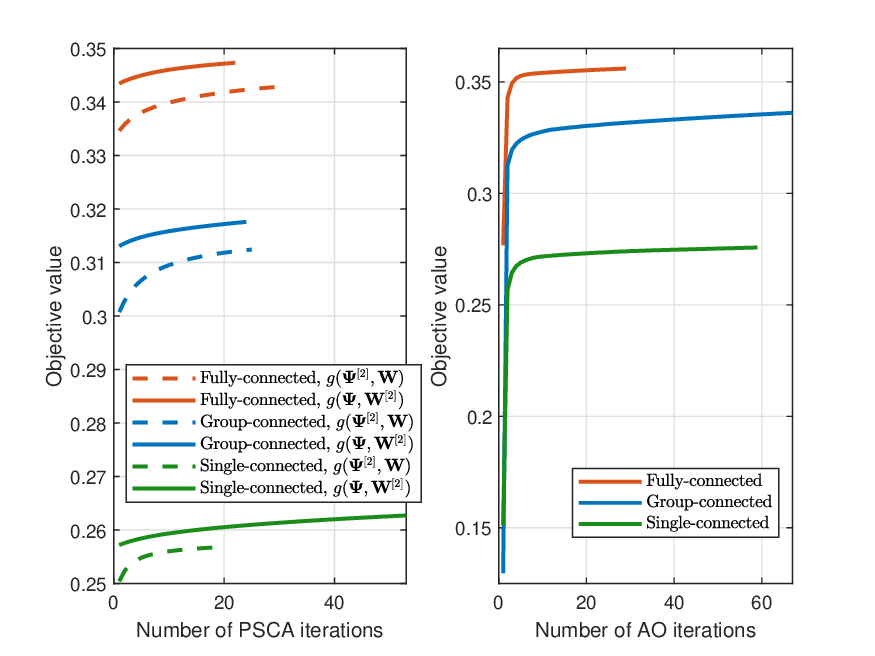}
    \caption{The convergence behavior of Algorithm \ref{a:AO-SCA}.}
    \label{fig: convergence}
\end{figure}
\begin{figure}[t]
  \centering
  \subfloat[The communication and sensing metric values versus $P$.]{\includegraphics[width=0.9\linewidth]{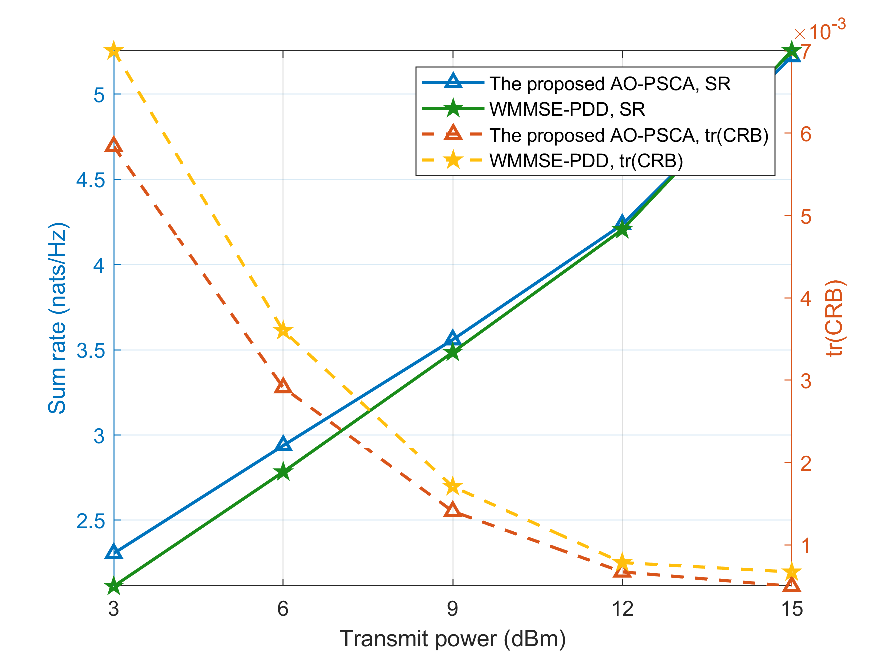}\label{fig:metricPt}}\\
  \subfloat[The average CPU time versus $P$.]{\includegraphics[width=0.9\linewidth]{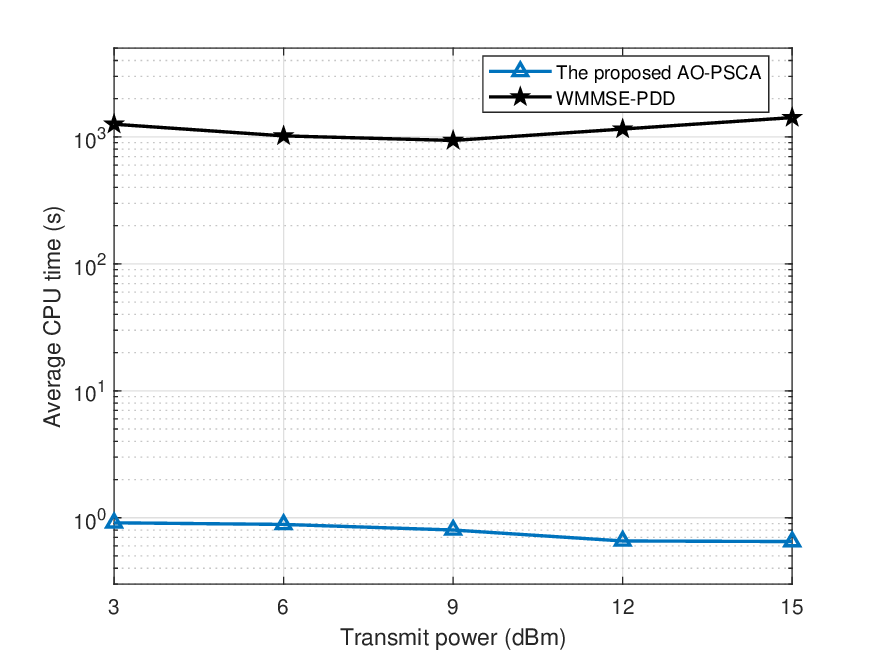}\label{fig:CPUtime}}
  \caption{Variation of metric values and average CPU time with transmit power ($N_I=8$, $K=2$, $Q=2$).}
  \label{fig:pt}
\end{figure}
To validate the efficacy of the proposed AO algorithm for BD-RIS-aided ISAC, we adopt a WMMSE-PDD iterative algorithm \cite{yzh_wmmsepdd} implemented via the CVX toolbox as a baseline.
For both algorithms, the convergence tolerance is set to $\tau = 10^{-3}$, and the initialization of the BD-RIS scattering matrix $\bm{\Psi}^{[0]}$ follows \cite{ckx24_transmitterBDRIS}, which maximizes the combined communication and sensing channel gains. While the active beamforming matrix $\mathbf{W} \triangleq [\mathbf{W}_c, \mathbf{W}_s]$ is initialized by setting $\mathbf{W}_c^{[0]}$ via maximum ratio transmission (MRT) and $\mathbf{W}_s^{[0]}$ via random Gaussian generation, followed by a normalization to satisfy the total power constraint, i.e., $\mathbf{W}^{[0]}=\bm{\Pi}_{\mathcal{W}}([\rho\bm{\Pi}_{\mathcal{W}}(\mathbf{W}_c^{[0]}), (1-\rho)\bm{\Pi}_{\mathcal{W}}(\mathbf{W}_s^{[0]})])$.
Furthermore, with the proposed AO method, we investigate three BD-RIS architectures, i.e., fully-, group- (with $G=4$), and single-connected BD-RIS. Note that the single-connected topology corresponds to the conventional diagonal RIS. 
All following numerical results are averaged over 100 random channel realizations.
\par
Fig. \ref{fig: convergence} illustrates the convergence behavior of the proposed AO-PSCA algorithm under different BD-RIS architectures, i.e., fully/group/single-connected BD-RIS.
Specifically, the left figure shows the convergence of subproblems with respect to $\mathbf{W}$ and $\bm{\Psi}$ at a specified AO iteration, while the right one presents the overall convergence of the AO algorithm.
It is observed that for all architectures, the algorithm exhibits a monotonic increase in the objective value until convergence, in accordance with the convergence analysis established in subsection \ref{sec:3_D}.
The overall algorithm converges within 70 iterations, thereby leading to fast convergence. 
\par
In Fig. \ref{fig:pt}, we compare the dual-functional performance and the average CPU time of the proposed AO method against the WMMSE-PDD baseline, where $K=2$ communication users and $N_I=8$ fully-connected BD-RIS elements are considered with transmit powers ranging from 3 dBm to 15 dBm.
Fig. \ref{fig:pt}(a) shows that the proposed AO algorithm yields better communication performance, i.e., higher SR, and sensing performance, i.e., lower trace of CRB across most transmit powers when compared to the WMMSE-PDD baseline, except for a slightly lower SR at 15 dBm.
Moreover, Fig. \ref{fig:pt}(b) shows that the proposed AO method significantly reduces the average CPU time by nearly three orders compared to the WMMSE-PDD baseline, which is attributed to its iterative closed-form solutions for each AO subproblem.
This thereby showcases its potential for future large-scale BD-RIS-assisted ISAC systems.
\par
In Fig. \ref{fig: tradeoff}, we present the trade-offs of fully/single-connected BD-RIS-assisted ISAC by varying the weighting factor $\rho$, where $Q=1,2,3$ sensing targets are considered.
As observed, the fully-connected BD-RIS-enabled ISAC system consistently exhibits a notable trade-off gain over its single-connected counterpart across different numbers of targets. 
This improvement is attributed to the extra design flexibility enabled by the full scattering matrix inherent in fully-connected BD-RIS.
With a growing number of targets, the trade-off regions of both architectures deteriorate due to the reduced power allocation per target.
Notably, the fully-connected BD-RIS-assisted ISAC system is capable of sensing more targets while maintaining the same SR performance for communication users.
This therefore highlights the potential of fully-connected BD-RIS to deliver superior sensing capabilities in ISAC systems.
\begin{figure}[t]
    \centering
    \includegraphics[width=0.9\linewidth]{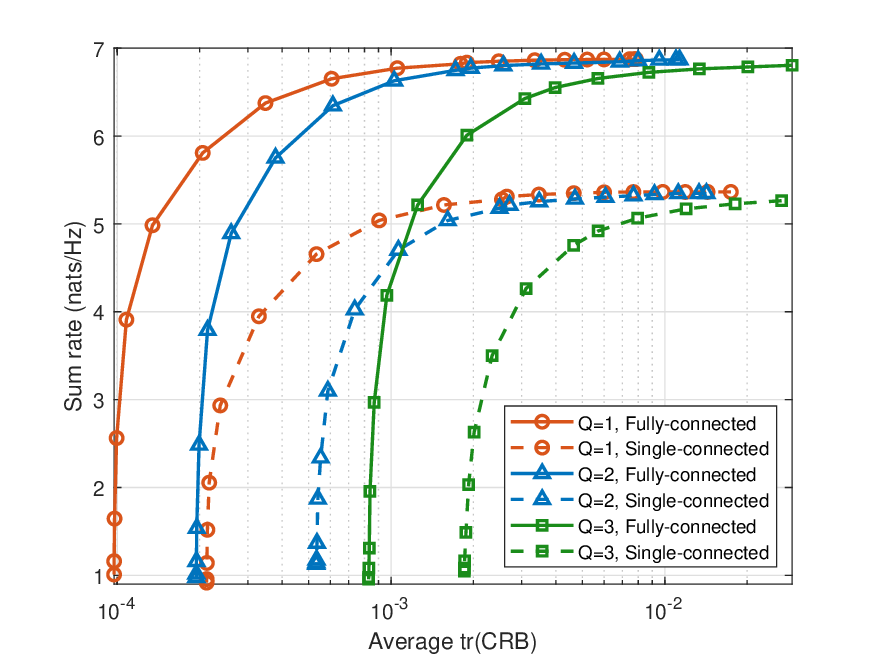}
    \caption{Illustration of the trade-off region for different numbers of targets ($N_I=32$, $K=4$).}
    \label{fig: tradeoff}
\end{figure}
\begin{figure}[t]
    \centering
    \includegraphics[width=0.9\linewidth]{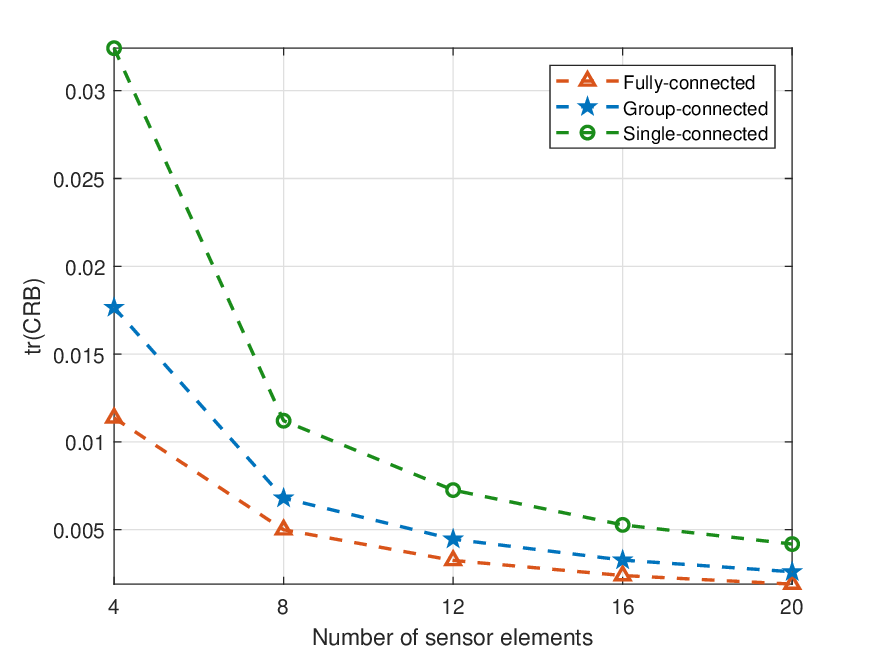}
    \caption{Variation of the sensing metric value with respect to the number of sensor elements ($N_I=32$, $K=4$, $Q=3$, $N_T=8$).}
    \label{fig: NS}
\end{figure}
\par
Fig. \ref{fig: NS} illustrates the impact of varying numbers of sensor elements on the trace of the sensing CRB, where $Q=3$ sensing targets and $N_T=8$ transmit antennas are considered.
It can be seen that across all three BD-RIS architectures, deploying more sensor elements leads to better estimation accuracy, i.e., lower trace of CRB.
This stems from the improved signal reception capabilities facilitated by the increased sensor elements.
Moreover, by leveraging its inherent full-matrix structure to construct sensing beams, the fully-connected BD-RIS-assisted ISAC achieves superior sensing accuracy compared to other architectures, highlighting its promise for advanced ISAC applications.
\begin{figure}[t]
    \centering
    \includegraphics[width=0.9\linewidth]{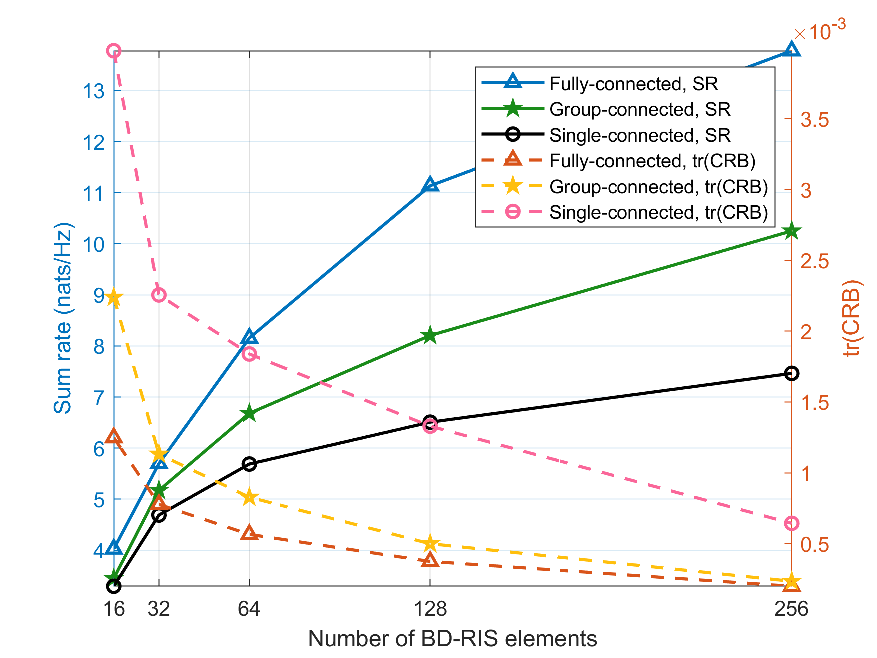}
    \caption{Variation of communication and sensing metric values with respect to the number of BD-RIS elements ($K=4$, $Q=2$).}
    \label{fig: RIS}
\end{figure}
\par
The communication SR and the trace of the sensing CRB across varying numbers of BD-RIS elements $N_I$ are depicted in Fig. \ref{fig: RIS}.
We observe that as the number of BD-RIS elements increases, better communication performance, i.e., higher SR, and sensing performance, i.e., lower trace of CRB, are achieved in all three architectures, thanks to the enlarged dimension of the BD-RIS scattering matrix determined by the number of elements.
The fully-connected BD-RIS-assisted ISAC consistently outperforms its group- and single-connected counterparts by offering additional DoFs through flexible inter-element connections, enabling more effective manipulation of wireless signals and joint optimization of communication and sensing.
Interestingly, the system aided by fully-connected BD-RIS achieves communication SR and trace of the sensing CRB comparable to those of group/single-connected BD-RIS-assisted systems with substantially fewer BD-RIS elements—for example, 64 elements versus roughly 128 and 256 elements for the group- and single-connected ones, respectively.
This thus demonstrates the superior capability of the fully-connected architecture to effectively exploit the available BD-RIS elements.
\endgroup
\section{Conclusion}\label{sec:5_conclu}
In this work, we propose a novel transmitter-side BD-RIS-enabled ISAC framework to alleviate the hardware burden caused by excessive RF chains and boost the sensing and communication performance.
The BD-RIS scattering matrix and the active beamforming matrix are optimized to jointly minimize the trace of the sensing CRB and maximize the communication SR.
To tackle the non-convexity stemming from the strong coupling of variables and BD-RIS constraints, we propose an efficient AO-PSCA algorithm, yielding iterative closed-form solutions.
Numerical results validate the superiority of the transmitter-side BD-RIS-assisted ISAC over its diagonal RIS-enabled counterpart in both communication and sensing performance. 
Furthermore, the proposed AO method serves as a low-complexity and practical algorithm with dual-functional performance comparable to that of the classic iterative algorithm.
Future research can extend to dynamic scenarios involving moving targets and focus on designing more efficient and universal algorithms applicable to other BD-RIS architectures, such as the recently proposed stem-connected BD-RIS \cite{zxh24_BD-RIS}.
\section*{Appendix}
\renewcommand{\thesubsection}{A}
\subsection[The derivations of the FIM]{The derivations of the FIM in \eqref{eq:FIM}} \label{appendix A}
Recall that $\mathbf{F}$ is the block FIM with respect to the parameter set $\bm{\xi }\triangleq \left \{ \bm{\theta} ,\bm{\phi}, \mathrm{Re}(\bm{\alpha}), \mathrm{Im}(\bm{\alpha}) \right \}^{T}=\{\bm{\xi}^1,\bm{\xi}^2,\bm{\xi}^3,\bm{\xi}^4\}^T$. 
Since the echo signal $\mathbf{y}_{s}[m]$ is a Gaussian observation following $\mathcal{CN}(\mathbf{v}_s[m],\sigma _{s}^{2}\mathbf{I}_{N_{S}})$ with $\mathbf{v}_s[m]=\mathbf{y}_{s}[m]-\mathbf{z}_{s}[m]$, each block in $\mathbf{F}$ can be calculated as \cite{lj08_CRB}
\begin{equation}\label{eq:FIM def}
\mathbf{F}_{\bm{\xi}^l\bm{\xi}^p}=\frac{2}{\sigma_s^2}\mathrm{Re}\left[\mathrm{tr}\left\{ \sum_{m=1}^M \frac{\partial\mathbf{v}_s^{H}[m] }{\partial\bm{\xi}^l }\frac{\partial\mathbf{v}_s[m]}{\partial\bm{\xi}^p}\right \}\right], \forall l,p.
\end{equation}
We take the derivation of $\mathbf{F}_{\bm{\xi}^1\bm{\xi}^1}$, i.e., $\mathbf{F}_{\bm{\theta}\bm{\theta}}$ as an example. To be specific, the partial derivative with respect to $\theta_i,\forall i \in\mathcal{Q}$ is given by
\begin{equation}
		\frac{\partial \mathbf{v}_s[m]}{\partial \theta_i }=\dot{\mathbf{B}}_{\bm{\theta}}\mathbf{e}_i\mathbf{e}_i^T\mathbf{U}\mathbf{A}^T\bm{\Psi}\mathbf{H}\mathbf{x}[m]+\mathbf{B}\mathbf{U}\mathbf{e}_i\mathbf{e}_i^T\dot{\mathbf{A}}_{\bm{\theta}}^T\bm{\Psi}\mathbf{H}\mathbf{x}[m],\forall i.
\end{equation}
where $\mathbf{e}_{i}$ refers to the $i$-th column of $\mathbf{I}_{Q}$. Based on the property $\mathrm{tr}(\mathbf{X}\mathbf{Y})=\mathrm{tr}(\mathbf{Y}\mathbf{X})$, we obtain
\begin{equation}
	\begin{aligned}
		&\mathbf{F}_{\theta_i\theta_j}=\frac{2}{\sigma_s^2}\mathrm{Re}\left[\mathrm{tr}\left\{ \sum_{m=1}^M \frac{\partial\mathbf{v}_s^{H}[m] }{\partial\theta_i }\frac{\partial\mathbf{v}_s[m]}{\partial\theta_j}\right \}\right]
        \\
        &=\frac{2M}{\sigma_s^2}\mathrm{Re}\Big (\mathbf{e}_i^T\dot{\mathbf{B}}_{\bm{\theta}}^H\dot{\mathbf{B}}_{\bm{\theta}}\mathbf{e}_j\mathbf{e}_j^T\mathbf{U}\mathbf{A}^T\bm{\Psi}\mathbf{H}\mathbf{R}_x\mathbf{H}^H\bm{\Psi}^H\mathbf{A}^\ast\mathbf{U}^H\mathbf{e}_i\\
        &+\mathbf{e}_i^T\dot{\mathbf{B}}_{\bm{\theta}}^H\mathbf{B} \mathbf{e}_j\mathbf{e}_j^T\mathbf{U}\dot{\mathbf{A}}_{\bm{\theta}}^{T}\bm{\Psi}\mathbf{H}\mathbf{R}_{x}\mathbf{H}^H\bm{\Psi}^H\mathbf{A}^\ast\mathbf{U}^H\mathbf{e}_i\\
        &+\mathbf{e}_i^T\mathbf{B}^{H}\dot{\mathbf{B}}_{\bm{\theta}}\mathbf{e}_j\mathbf{e}_j^T\mathbf{U}\mathbf{A}^{T}\bm{\Psi}\mathbf{H}\mathbf{R}_{x}\mathbf{H}^H\bm{\Psi}^H\dot{\mathbf{A}}_{\bm{\theta}}^\ast\mathbf{U}^H\mathbf{e}_i\\
        &+\mathbf{e}_i^T\mathbf{B}^{H}\mathbf{B}\mathbf{e}_j\mathbf{e}_j^T\mathbf{U}\dot{\mathbf{A}}_{\bm{\theta}}^{T}\bm{\Psi}\mathbf{H}\mathbf{R}_{x}\mathbf{H}^H\bm{\Psi}^H\dot{\mathbf{A}}_{\bm{\theta}}^\ast\mathbf{U}^H\mathbf{e}_i \Big) \\
		&=\frac{2M}{\sigma_s^2}\mathrm{Re}\Big [ (\dot{\mathbf{B}}_{\bm{\theta}}^H\dot{\mathbf{B}}_{\bm{\theta}})_{ij}(\mathbf{U}\mathbf{A}^T\bm{\Psi}\mathbf{H}\mathbf{R}_x\mathbf{H}^H\bm{\Psi}^H\mathbf{A}^\ast\mathbf{U}^H)_{ji}\\
        &+(\dot{\mathbf{B}}_{\bm{\theta}}^H\mathbf{B})_{ij} (\mathbf{U}\dot{\mathbf{A}}_{\bm{\theta}}^{T}\bm{\Psi}\mathbf{H}\mathbf{R}_{x}\mathbf{H}^H\bm{\Psi}^H\mathbf{A}^\ast\mathbf{U}^H)_{ji}\\
        &+(\mathbf{B}^{H}\dot{\mathbf{B}}_{\bm{\theta}})_{ij} (\mathbf{U}\mathbf{A}^{T}\bm{\Psi}\mathbf{H}\mathbf{R}_{x}\mathbf{H}^H\bm{\Psi}^H\dot{\mathbf{A}}_{\bm{\theta}}^\ast\mathbf{U}^H)_{ji}\\
        &+(\mathbf{B}^{H}\mathbf{B})_{ij} (\mathbf{U}\dot{\mathbf{A}}_{\bm{\theta}}^{T}\bm{\Psi}\mathbf{H}\mathbf{R}_{x}\mathbf{H}^H\bm{\Psi}^H\dot{\mathbf{A}}_{\bm{\theta}}^\ast\mathbf{U}^H)_{ji}\Big], \forall i,j\in\mathcal{Q},
	\end{aligned}
 \end{equation}
where $(\cdot )_{ij}$ denotes the element at the $i$-th row and the $j$-th column. Obviously, we have $\mathbf{F}_{\theta_i\theta_j}=\frac{2M}{\sigma_s^2}\mathrm{Re}([\mathbf{F}_{11}]_{ij})$, i.e., $\mathbf{F}_{\bm{\theta}\bm{\theta}}=\frac{2M}{\sigma_s^2}\mathrm{Re}(\mathbf{F}_{11})$, with $\mathbf{F}_{11}$ specified in \eqref{eq:FIM entry}. Other terms in the FIM $\mathbf{F}$ can be derived following the same approach.
Hence, we obtain the FIM in \eqref{eq:FIM}.
\renewcommand{\thesubsection}{B}
\subsection[The derivations of the surrogate function]{The derivations of the surrogate function in \eqref{eq:f_psi_S2}}\label{appendix B}
Thanks to the symmetric blockwise structures of $\mathbf{F}$ and $\mathbf{J}$, the term $\mathrm{tr}(\mathbf{J}^{[l]}\mathbf{F})$ of the original surrogate function at the $l$-th iteration in \eqref{eq:f_psi_S1} can be expanded and rewritten as
\begin{equation}\label{eq:appB_trJF}
\begin{aligned}
    &\mathrm{tr}(\mathbf{J}^{[l]}\mathbf{F})=\frac{2M}{\sigma_s^2}\mathrm{Re}\bigg[\mathrm{tr}\Big\{\mathbf{F}_{11}(\mathbf{J}_{11}^{[l]})^T+2\mathbf{F}_{12}(\mathbf{J}_{12}^{[l]})^T\\
    &+2\mathbf{F}_{13}(\mathbf{J}_{13}^{[l]}+j\mathbf{J}_{14}^{[l]})^T+\mathbf{F}_{22}(\mathbf{J}_{22}^{[l]})^T+2\mathbf{F}_{23}(\mathbf{J}_{23}^{[l]}+j\mathbf{J}_{24}^{[l]})^T\\
    &+\mathbf{F}_{33}(\mathbf{J}_{33}^{[l]}+\mathbf{J}_{44}^{[l]}+2j\mathbf{J}_{34}^{[l]})^T\Big\}\bigg],
\end{aligned}
\end{equation}
where we applied the property $\mathrm{tr}(\mathbf{X}\mathbf{Y}^T)=\mathrm{tr}(\mathbf{X}^T\mathbf{Y})$. 
With another property $\mathrm{tr}(\mathbf{X}^T(\mathbf{Y}^T\odot\mathbf{Z}))=\mathrm{tr}(\mathbf{Y}(\mathbf{X}\odot\mathbf{Z}))$, the first term in \eqref{eq:appB_trJF}, i.e., $\mathrm{Re}\{\mathrm{tr}(\mathbf{F}_{11}(\mathbf{J}_{11}^{[l]})^T)\}$ can be further reformulated as
\begin{equation}
\begin{aligned}
        &\mathrm{Re}\bigg[\mathrm{tr}\Big\{\Big((\dot{\mathbf{B}}_{\bm{\theta}}^H \dot{\mathbf{B}}_{\bm{\theta}})\odot (\mathbf{U}\mathbf{A}^{T}\bm{\Psi}\mathbf{H}\mathbf{R}_{x}\mathbf{H}^H\bm{\Psi}^H\mathbf{A}^\ast\mathbf{U}^H)^T\\&+(\dot{\mathbf{B}}_{\bm{\theta}}^H\mathbf{B})\odot (\mathbf{U}\dot{\mathbf{A}}_{\bm{\theta}}^{T}\bm{\Psi}\mathbf{H}\mathbf{R}_{x}\mathbf{H}^H\bm{\Psi}^H\mathbf{A}^\ast\mathbf{U}^H)^T\\&+(\mathbf{B}^{H}\dot{\mathbf{B}}_{\bm{\theta}})\odot (\mathbf{U}\mathbf{A}^{T}\bm{\Psi}\mathbf{H}\mathbf{R}_{x}\mathbf{H}^H\bm{\Psi}^H\dot{\mathbf{A}}_{\bm{\theta}}^\ast\mathbf{U}^H)^T\\&+(\mathbf{B}^{H}\mathbf{B})\odot (\mathbf{U}\dot{\mathbf{A}}_{\bm{\theta}}^{T}\bm{\Psi}\mathbf{H}\mathbf{R}_{x}\mathbf{H}^H\bm{\Psi}^H\dot{\mathbf{A}}_{\bm{\theta}}^\ast\mathbf{U}^H)^T\Big)(\mathbf{J}_{11}^{[l]})^T\Big\}\bigg]\\
        &=\mathrm{Re}\bigg[\mathrm{tr}\Big\{\bm{\Psi}\mathbf{H}\mathbf{R}_{x}\mathbf{H}^H\bm{\Psi}^H\mathbf{A}^\ast\mathbf{U}^H\big((\dot{\mathbf{B}}_{\bm{\theta}}^H \dot{\mathbf{B}}_{\bm{\theta}})\odot \mathbf{J}_{11}^{[l]}\big)\mathbf{U}\mathbf{A}^{T}\\&+\bm{\Psi}\mathbf{H}\mathbf{R}_{x}\mathbf{H}^H\bm{\Psi}^H\mathbf{A}^\ast\mathbf{U}^H\big((\dot{\mathbf{B}}_{\bm{\theta}}^H\mathbf{B})\odot\mathbf{J}_{11}^{[l]} \big)\mathbf{U}\dot{\mathbf{A}}_{\bm{\theta}}^{T}\\&+\bm{\Psi}\mathbf{H}\mathbf{R}_{x}\mathbf{H}^H\bm{\Psi}^H\dot{\mathbf{A}}_{\bm{\theta}}^\ast\mathbf{U}^H\big((\mathbf{B}^{H}\dot{\mathbf{B}}_{\bm{\theta}})\odot\mathbf{J}_{11}^{[l]} \big)\mathbf{U}\mathbf{A}^{T}\\&+\bm{\Psi}\mathbf{H}\mathbf{R}_{x}\mathbf{H}^H\bm{\Psi}^H\dot{\mathbf{A}}_{\bm{\theta}}^\ast\mathbf{U}^H\big((\mathbf{B}^{H}\mathbf{B})\odot \mathbf{J}_{11}^{[l]}\big)\mathbf{U}\dot{\mathbf{A}}_{\bm{\theta}}^{T}\Big\}\bigg]\\
        &=\mathrm{Re}\left\{\mathrm{tr}\left(\bm{\Psi}\mathbf{H}\mathbf{R}_x\mathbf{H}^H\bm{\Psi}^H\mathbf{\Sigma}_{11}^{[l]}\right)\right\},
\end{aligned}
\end{equation}
where $\mathbf{\Sigma}_{11}^{[l]}$ is defined in \eqref{eq:Sigma}. The remaining terms can be rewritten following the same way with other auxiliary matrices given by \eqref{eq:Sigma}. Hence, we obtain the surrogate function in \eqref{eq:f_psi_S2}.
\renewcommand{\thesubsection}{C}
\subsection[Proof of Proposition 2]{Proof of Proposition \ref{prop:2}}\label{appendix C}
By introducing auxiliary variables $T_k=\|\mathbf{h}_{k}^H\bm{\Psi}\mathbf{H}\mathbf{W}\|_2^2+\sigma_c^2$, $S_k=\left|\mathbf{h}_{k}^H\bm{\Psi}\mathbf{H}\mathbf{w}_{k}\right|^2$ and $I_k=T_k-S_k, \forall k$, we compute the gradients of the communication and sensing objective functions, i.e., $g_c(\bm{\Psi})$, $g_s(\bm{\Psi})$ given in \eqref{eq:g_h} as follows.
\par
The gradient of $g_c(\bm{\Psi})$ with respect to $\bm{\Psi}$ is expressed as
\begin{equation}\label{eq:gradient_g_c}
\begin{aligned}
    \frac{\partial g_c(\bm{\Psi})}{\partial \bm{\Psi}}&=\frac{\rho}{V_c}\sum_{k\in\mathcal{K}}\frac{\partial \log(1+\gamma_k)}{\partial \bm{\Psi}}=\frac{\rho}{V_c}\sum_{k\in\mathcal{K}}\frac{\partial \log(1+S_k/I_k)}{\partial \bm{\Psi}}\\
    &=\frac{2\rho}{V_c}\sum_{k\in\mathcal{K}}\left(1+\frac{S_k}{I_k}\right)^{-1}\bigg(\frac{I_k\mathbf{h}_k\mathbf{h}_k^H\bm{\Psi}\mathbf{H}\mathbf{w}_k\mathbf{w}_k^H\mathbf{H}^H}{I_k^2}\\
    &-\frac{S_k\mathbf{h}_k\mathbf{h}_k^H\bm{\Psi}\mathbf{C}-S_k\mathbf{h}_k\mathbf{h}_k^H\bm{\Psi}\mathbf{H}\mathbf{w}_k\mathbf{w}_k^H\mathbf{H}^H}{I_k^2}\bigg)\\
    &=\frac{2\rho}{V_c}\sum_{k\in\mathcal{K}}\left(\zeta_k^\ast\mathbf{h}_k\mathbf{w}_k^H\mathbf{H}^H-\eta_k\mathbf{h}_k\mathbf{h}_k^H\bm{\Psi}\mathbf{C}\right)\\
    &=\frac{2\rho}{V_c}\mathbf{H}_c\mathbf{E}_1^H\mathbf{W}_c^H\mathbf{H}^H-\frac{2\rho}{V_c}\mathbf{H}_c\mathbf{E}_2\mathbf{H}_c^H\bm{\Psi}\mathbf{C},
\end{aligned}
\end{equation}
where we use the equality $(1+\frac{S_k}{I_k})^{-1}=\frac{I_k}{T_k}$, and $\mathbf{H}_c$, $\mathbf{C}$, $\mathbf{E}_1$, $\mathbf{E}_2$ are matrices defined in \eqref{eq:H_C_E}. Note that with $\bm{\Psi}=\bm{\Psi}^{[l]}$, the gradient in \eqref{eq:gradient_g_c} is identical to that of $h_c^{[l]}(\bm{\Psi})$.
\par
The gradient of the sensing objective function $g_s(\bm{\Psi})$ is expressed as
\begin{equation}
\scalebox{0.95}{$
    \begin{aligned}\label{eq:partial_gs}
        \frac{\partial g_s(\bm{\Psi})}{\partial \bm{\Psi}}&=\frac{-(1-\rho)}{V_s}\frac{\partial \mathrm{tr}(\mathbf{F}^{-1})}{\partial \bm{\Psi}}\\
        &=\frac{-(1-\rho)}{V_s}\sum_{l,p=1}^4 \sum_{i,j=1}^Q \frac{\partial \mathrm{tr}(\mathbf{F}^{-1})}{\partial (\mathbf{F}_{\bm{\xi}^l\bm{\xi}^p})_{ij}}\frac{\partial (\mathbf{F}_{\bm{\xi}^l\bm{\xi}^p})_{ij}}{\partial \bm{\Psi}}\\
        &=\frac{1-\rho}{V_s}\sum_{l,p=1}^4 \sum_{i,j=1}^Q (\mathbf{J}_{lp})_{ij}\frac{\partial (\mathbf{F}_{\bm{\xi}^l\bm{\xi}^p})_{ij}}{\partial \bm{\Psi}},
    \end{aligned}
    $}
\end{equation}
where we use the gradient $-\partial \mathrm{tr}(\mathbf{F}^{-1})/\partial \mathbf{F}=\mathbf{F}^{-2}=\mathbf{J}$ with $\mathbf{J}$ given by \eqref{eq:J}, and $\mathbf{F}_{\bm{\xi}^l\bm{\xi}^p}, \forall l,p$ is defined in \eqref{eq:FIM def}. We take the derivation of $l=1, p=1$ in \eqref{eq:partial_gs} as an example, which is calculated as
\begin{equation}
    \scalebox{0.93}{$
    \begin{aligned}
        &\frac{1-\rho}{V_s}\sum_{i,j=1}^Q(\mathbf{J}_{11})_{ij}\frac{\partial \frac{2M}{\sigma_s^2}\mathrm{Re}(\mathbf{F}_{11})_{ij}}{\partial \bm{\Psi}}\\
        &=\frac{1-\rho}{V_s}\frac{4M}{\sigma_s^2}\sum_{i,j=1}^Q(\mathbf{J}_{11})_{ij}\mathrm{Re}\Big\{\mathbf{A}^\ast\mathbf{U}^H\mathbf{e}_i\mathbf{e}_i^T\dot{\mathbf{B}}_{\bm{\theta}}^H \dot{\mathbf{B}}_{\bm{\theta}}\mathbf{e}_j\mathbf{e}_j^T\mathbf{U}\mathbf{A}^{T}\\
        &+\mathbf{A}^\ast\mathbf{U}^H\mathbf{e}_i\mathbf{e}_i^T\dot{\mathbf{B}}_{\bm{\theta}}^H\mathbf{B} \mathbf{e}_j\mathbf{e}_j^T\mathbf{U}\dot{\mathbf{A}}_{\bm{\theta}}^{T}
        +\dot{\mathbf{A}}_{\bm{\theta}}^\ast\mathbf{U}^H\mathbf{e}_i\mathbf{e}_i^T\mathbf{B}^{H}\dot{\mathbf{B}}_{\bm{\theta}} \mathbf{e}_j\mathbf{e}_j^T\mathbf{U}\mathbf{A}^{T}\\
        &+\dot{\mathbf{A}}_{\bm{\theta}}^\ast\mathbf{U}^H\mathbf{e}_i\mathbf{e}_i^T\mathbf{B}^{H}\mathbf{B}\mathbf{e}_j\mathbf{e}_j^T\mathbf{U}\dot{\mathbf{A}}_{\bm{\theta}}^{T}
        \Big\}\bm{\Psi}\mathbf{C}\\
        &=\frac{1-\rho}{V_s}\frac{4M}{\sigma_s^2}\mathrm{Re}\Big\{\mathbf{A}^\ast\mathbf{U}^H\big((\dot{\mathbf{B}}_{\bm{\theta}}^H \dot{\mathbf{B}}_{\bm{\theta}})\odot\mathbf{J}_{11}\big)\mathbf{U}\mathbf{A}^{T}\\
        &+\mathbf{A}^\ast\mathbf{U}^H\big((\dot{\mathbf{B}}_{\bm{\theta}}^H\mathbf{B})\odot\mathbf{J}_{11}\big)\mathbf{U}\dot{\mathbf{A}}_{\bm{\theta}}^{T}
        +\dot{\mathbf{A}}_{\bm{\theta}}^\ast\mathbf{U}^H\big((\mathbf{B}^{H}\dot{\mathbf{B}}_{\bm{\theta}})\odot\mathbf{J}_{11}\big)\mathbf{U}\mathbf{A}^{T}\\
        &+\dot{\mathbf{A}}_{\bm{\theta}}^\ast\mathbf{U}^H\big((\mathbf{B}^{H}\mathbf{B})\odot\mathbf{J}_{11}\big)\mathbf{U}\dot{\mathbf{A}}_{\bm{\theta}}^{T}
        \Big\}\bm{\Psi}\mathbf{C}\\
        &=\frac{1-\rho}{V_s}\frac{2M}{\sigma_s^2}(\bm{\Sigma}_{11}+\bm{\Sigma}_{11}^H)\bm{\Psi}\mathbf{C}.
    \end{aligned}$}
\end{equation}
Similarly, based on $\mathbf{F}_{lp}$ and $\bm{\Sigma}_{lp}$ given by \eqref{eq:FIM entry} and \eqref{eq:Sigma}, we obtain the gradient as $\partial g_s(\bm{\Psi})/\partial \bm{\Psi}=(1-\rho)/V_s(\bm{\Sigma}+\bm{\Sigma}^H)\bm{\Psi}\mathbf{C}$, where $\bm{\Sigma}$ is specified in \eqref{eq:Sigma}.
We observe that when $\bm{\Psi}=\bm{\Psi}^{[l]}$, this gradient takes the same form as that of $h_s^{[l]}(\bm{\Psi})$. Hence, we obtain \eqref{eq:grad} by combining the established equalities for the communication and sensing part, i.e., $\nabla g_c(\bm{\Psi})|_{\bm{\Psi}=\bm{\Psi}^{[l]}}=\nabla h_c^{[l]}(\bm{\Psi})$ and $\nabla g_s(\bm{\Psi})|_{\bm{\Psi}=\bm{\Psi}^{[l]}}
        =\nabla h_s^{[l]}(\bm{\Psi})$.
\section*{Acknowledgement}
The authors thank Tianyu Fang at the University of Oulu for valuable discussions.
\bibliographystyle{IEEEtran}  
\bibliography{ref}

\begin{thebibliography}{10}
\providecommand{\url}[1]{#1}
\csname url@samestyle\endcsname
\providecommand{\newblock}{\relax}
\providecommand{\bibinfo}[2]{#2}
\providecommand{\BIBentrySTDinterwordspacing}{\spaceskip=0pt\relax}
\providecommand{\BIBentryALTinterwordstretchfactor}{4}
\providecommand{\BIBentryALTinterwordspacing}{\spaceskip=\fontdimen2\font plus
\BIBentryALTinterwordstretchfactor\fontdimen3\font minus \fontdimen4\font\relax}
\providecommand{\BIBforeignlanguage}[2]{{%
\expandafter\ifx\csname l@#1\endcsname\relax
\typeout{** WARNING: IEEEtran.bst: No hyphenation pattern has been}%
\typeout{** loaded for the language `#1'. Using the pattern for}%
\typeout{** the default language instead.}%
\else
\language=\csname l@#1\endcsname
\fi
#2}}
\providecommand{\BIBdecl}{\relax}
\BIBdecl

\bibitem{ckx24_transmitterBDRIS}
K.~Chen \emph{et~al.}, ``Transmitter side beyond-diagonal {RIS} for {mmWave} integrated sensing and communications,'' in \emph{Proc. IEEE Int. Workshop Signal Process. Adv. Wireless Commun. (SPAWC)}, 2024, pp. 951--955.

\bibitem{lf22_tutorial}
F.~Liu \emph{et~al.}, ``Integrated sensing and communications: Toward dual-functional wireless networks for {6G} and beyond,'' \emph{IEEE J. Sel. Areas Commun.}, vol.~40, no.~6, pp. 1728--1767, 2022.

\bibitem{kaushik24_ISAC}
A.~Kaushik \emph{et~al.}, ``Toward integrated sensing and communications for {6G}: Key enabling technologies, standardization, and challenges,'' \emph{IEEE Commun. Stand. Mag.}, vol.~8, no.~2, pp. 52--59, 2024.

\bibitem{lf20_ISAC}
F.~Liu \emph{et~al.}, ``Joint radar and communication design: Applications, state-of-the-art, and the road ahead,'' \emph{IEEE Trans. Commun.}, vol.~68, no.~6, pp. 3834--3862, 2020.

\bibitem{lsh24_ISACtutorial}
S.~Lu \emph{et~al.}, ``Integrated sensing and communications: Recent advances and ten open challenges,'' \emph{IEEE Internet Things J.}, vol.~11, no.~11, pp. 19\,094--19\,120, 2024.

\bibitem{lr23_RISISAC}
R.~Liu \emph{et~al.}, ``Integrated sensing and communication with reconfigurable intelligent surfaces: Opportunities, applications, and future directions,'' \emph{IEEE Wireless Commun.}, vol.~30, no.~1, pp. 50--57, 2023.

\bibitem{lr23_RISISAC2}
------, ``{SNR/CRB}-constrained joint beamforming and reflection designs for {RIS-ISAC} systems,'' \emph{IEEE Trans. Wireless Commun.}, vol.~23, no.~7, pp. 7456--7470, 2023.

\bibitem{wzl23_STARS}
Z.~Wang \emph{et~al.}, ``{STARS} enabled integrated sensing and communications,'' \emph{IEEE Trans. Wireless Commun.}, vol.~22, no.~10, pp. 6750--6765, 2023.

\bibitem{wwj24_normalized}
W.~Wei \emph{et~al.}, ``{Cram{\'e}r}-rao bound and secure transmission trade-off design for semi-{IRS}-enabled {ISAC},'' \emph{IEEE Trans. Wireless Commun.}, vol.~23, no.~11, pp. 15\,753--15\,767, 2024.

\bibitem{ly24_mmWaveISAC}
W.~Lyu \emph{et~al.}, ``{CRB} minimization for {RIS}-aided {mmWave} integrated sensing and communications,'' \emph{IEEE Internet Things J.}, vol.~11, no.~10, pp. 18\,381--18\,393, 2024.

\bibitem{wl25_RISmmWaveISAC}
L.~Wang \emph{et~al.}, ``Joint hybrid beamforming and {RIS} phase shift design for {RIS}-enabled {mmWave} {ISAC} system,'' \emph{IEEE Trans. Veh. Technol.}, vol.~74, no.~6, pp. 9149--9164, 2025.

\bibitem{jamali21_activeantenna}
V.~Jamali \emph{et~al.}, ``Intelligent surface-aided transmitter architectures for millimeter-wave ultra massive {MIMO} systems,'' \emph{IEEE Open J. Commun. Soc.}, vol.~2, pp. 144--167, 2021.

\bibitem{buzzi21_transmitRIS}
S.~Buzzi \emph{et~al.}, ``{RIS}-aided massive {MIMO}: Achieving large multiplexing gains with non-large arrays,'' in \emph{Int. Workshop on Smart Antennas (WSA)}, 2021, pp. 1--6.

\bibitem{dwn23_ITScommunication}
W.~Du \emph{et~al.}, ``Hybrid beamforming design for {ITS}-assisted wireless networks,'' \emph{IEEE Wireless Commun. Lett.}, vol.~12, no.~3, pp. 451--455, 2023.

\bibitem{ssp22_BDRIS}
S.~Shen \emph{et~al.}, ``Modeling and architecture design of reconfigurable intelligent surfaces using scattering parameter network analysis,'' \emph{IEEE Trans. Wireless Commun.}, vol.~21, no.~2, pp. 1229--1243, 2022.

\bibitem{lhy22_BDRIS}
H.~Li \emph{et~al.}, ``Beyond diagonal reconfigurable intelligent surfaces: From transmitting and reflecting modes to single-, group-, and fully-connected architectures,'' \emph{IEEE Trans. Wireless Commun.}, vol.~22, no.~4, pp. 2311--2324, 2022.

\bibitem{lhy25_BDRIStutorial}
------, ``A tutorial on beyond-diagonal reconfigurable intelligent surfaces: Modeling, architectures, system design and optimization, and applications,'' \emph{arXiv preprint arXiv:2505.16504}, 2025.

\bibitem{lzr24_ISAC_BDRIS}
Z.~Liu \emph{et~al.}, ``Enhancing {ISAC} network throughput using beyond diagonal {RIS},'' \emph{IEEE Wireless Commun. Lett.}, vol.~13, no.~6, pp. 1670--1674, 2024.

\bibitem{gzh24_BDRISISAC}
Z.~Guang \emph{et~al.}, ``Power minimization for {ISAC} system using beyond diagonal reconfigurable intelligent surface,'' \emph{IEEE Trans. Veh. Technol.}, vol.~73, no.~9, pp. 13\,950--13\,955, 2024.

\bibitem{esmaeil24_BDRISISAC}
T.~Esmaeilbeig \emph{et~al.}, ``Beyond diagonal {RIS}: Key to next-generation integrated sensing and communications?'' \emph{IEEE Signal Process Lett.}, vol.~32, pp. 216--220, 2025.

\bibitem{Nguyen25_BDRIS_ISAC}
T.~L. Nguyen \emph{et~al.}, ``Beyond diagonal {RIS} for {ISAC} network: Statistical analysis and network parameter estimation,'' \emph{arXiv preprint arXiv:2502.12916}, 2025.

\bibitem{pxl24_BDRIS_ISAC}
X.~Peng \emph{et~al.}, ``Beyond-diagonal {RIS} aided {DFRC} systems: A joint beamforming optimization design method,'' in \emph{Proc. IEEE Int. Conf. Commun. China (ICCC) Workshops}, 2024, pp. 782--787.

\bibitem{WBW23_BDRISISAC}
B.~Wang \emph{et~al.}, ``A dual-function radar-communication system empowered by beyond diagonal reconfigurable intelligent surface,'' \emph{IEEE Trans. Commun.}, vol.~73, no.~3, pp. 1501--1516, 2025.

\bibitem{zs25_BDRIS_ISAC}
S.~Zheng \emph{et~al.}, ``Beyond diagonal intelligent reflecting surface aided integrated sensing and communication,'' \emph{arXiv preprint arXiv:2505.16230}, 2025.

\bibitem{wdw25_BDRIS_ISAC}
D.~Wang \emph{et~al.}, ``Enhanced {ISAC} framework for moving target assisted by beyond-diagonal {RIS}: Accurate localization and efficient communication,'' \emph{IEEE Trans. Network Sci. Eng.}, vol.~12, no.~5, pp. 4299--4315, 2025.

\bibitem{zxq25_BDRISISAC}
X.~Zhang \emph{et~al.}, ``Optimizing rate-{CRB} performance for beyond diagonal reconfigurable intelligent surface enabled {ISAC},'' \emph{IEEE Commun. Lett.}, 2025.

\bibitem{lj08_CRB}
J.~Li \emph{et~al.}, ``Range compression and waveform optimization for {MIMO} radar: A cram\'er–rao bound based study,'' \emph{IEEE Trans. Signal Process.}, vol.~56, no.~1, pp. 218--232, 2008.

\bibitem{Mishra23_BDRIScomm}
A.~Mishra \emph{et~al.}, ``Transmitter side beyond-diagonal reconfigurable intelligent surface for massive {MIMO} networks,'' \emph{IEEE Wireless Commun. Lett.}, vol.~13, no.~2, pp. 352--356, 2024.

\bibitem{sxd22_sensingRIS}
X.~Shao \emph{et~al.}, ``Target sensing with intelligent reflecting surface: Architecture and performance,'' \emph{IEEE J. Sel. Areas Commun.}, vol.~40, no.~7, pp. 2070--2084, 2022.

\bibitem{fty25_ISAClowcomplexity}
T.~Fang \emph{et~al.}, ``Optimal {ISAC} beamforming structure and efficient algorithms for sum rate and {CRLB} balancing,'' \emph{arXiv preprint arXiv:2503.09489}, 2025.

\bibitem{fty23_Rk}
------, ``Optimal beamforming structure and efficient optimization algorithms for generalized multi-group multicast beamforming optimization,'' \emph{IEEE Trans. Signal Process.}, vol.~73, pp. 2719--2735, 2025.

\bibitem{sy16_MM}
Y.~Sun \emph{et~al.}, ``Majorization-minimization algorithms in signal processing, communications, and machine learning,'' \emph{IEEE Trans. Signal Process.}, vol.~65, no.~3, pp. 794--816, 2016.

\bibitem{zxh25_jointpassive}
X.~Zhou \emph{et~al.}, ``Joint active and passive beamforming optimization for beyond diagonal {RIS}-aided multi-user communications,'' \emph{IEEE Commun. Lett.}, vol.~29, no.~3, pp. 517--521, 2025.

\bibitem{fty24_BDRISlowcomplexity}
T.~Fang \emph{et~al.}, ``A low-complexity beamforming design for beyond-diagonal {RIS} aided multi-user networks,'' \emph{IEEE Commun. Lett.}, vol.~28, no.~1, pp. 203--207, 2024.

\bibitem{yzh_wmmsepdd}
Z.~Yao \emph{et~al.}, ``{RSMA} assisted {ISAC} with hybrid beamforming,'' in \emph{Proc. IEEE Wireless Commun. Netw. Conf. (WCNC)}, 2025, pp. 01--07.

\bibitem{zxh24_BD-RIS}
X.~Zhou \emph{et~al.}, ``A novel {Q-stem} connected architecture for beyond-diagonal reconfigurable intelligent surfaces,'' \emph{arXiv preprint arXiv:2411.18480}, 2024.

\end{thebibliography}

\end{document}